\documentclass[twocolumn,showpacs,preprintnumbers,amsmath,amssymb,superscriptaddress,pra,longbibliography]{revtex4-1}

\usepackage{graphicx}
\usepackage{dcolumn}
\usepackage{bm}
\usepackage{graphics}
\usepackage{subfigure}
\usepackage{graphicx}
\usepackage{epsfig}
\usepackage{epstopdf}
\usepackage{mathtools}
\usepackage{xcolor}

\allowdisplaybreaks

\begin{document}

\title{Quantum oscillations in Dirac magnetoplasmons}

\author{Johannes Hofmann}
\affiliation{Department of Applied Mathematics and Theoretical Physics, University of Cambridge, Centre for Mathematical Sciences, Cambridge CB3 0WA, United Kingdom}
\affiliation{TCM Group, Cavendish Laboratory, University of Cambridge, Cambridge CB3 0HE, United Kingdom}

\date{\today}

\begin{abstract}
The plasmon frequency in standard electron gases with a parabolic single-particle dispersion is a purely classical quantity that is not sensitive to electron interactions or the equation of state. We demonstrate that this canonical result no longer holds for plasmons in three-dimensional semimetals, which can thus be used to probe many-body effects in these systems. In particular, we show that the plasmon frequency in an external magnetic field displays quantum oscillations, which is not the case for the electron gas. Using the random phase approximation, results are presented for the magnetoplasmon dispersion and the loss function in Dirac semimetals. We include a full discussion of the loss function in a magnetic field as a function of the direction of propagation with respect to the magnetic field direction and discuss the transition from large magnetic fields to the low-field limit.
\end{abstract}

\maketitle

\section{Introduction}

Quasirelativistic phases of matter, such as graphene~\cite{castroneto09,dassarma11}, transition metal dichalcogenides~\cite{manzelli17}, topological insulators~\cite{hasan10}, or Dirac and Weyl semimetals~\cite{armitage18}, have been widely explored over the past decade~\cite{nature14}. Their Hamiltonian resembles that of relativistic Dirac particles (with the Fermi velocity or the band gap playing the role of the speed of light or the Dirac mass). The continued interest in theses systems is, to a considerable part, due to possible applications in plasmonics~\cite{wunsch06,hwang07,ju11,grigorenko12,dipeitro13}. In this paper, we point out that the Dirac plasmon frequency in three-dimensional semimetals (3DSM) is sensitive to the equation of state and, as one manifestation of this, shows quantum oscillations. By contrast, quantum oscillations or other interaction effects are absent for the electron gas (which describes solids with parabolic bands), where they only appear as higher-order corrections in the wavenumber ${\cal O}(q^2/\hbar^2)$~\cite{pines89}. Our main focus are three-dimensional gapless Dirac semimetals, for which the single-particle dispersion is a linear function of momentum. For these systems, the theoretical study of plasmons is a flourishing subfield~\cite{lv13,panfilov14,zhou15,hofmann15,pellegrino15,hofmann16,kotov16,song17,gorbar17a,gorbar17b,andolina18,kotov18,kozii18,chen19}, and first measurements of plasmons have recently been reported in optical studies~\cite{sushkov15,chen15,jenkins16,xu16,chanana18} and electron-loss spectroscopy~\cite{chiarello18}.

To put the results of this paper in context, we recall the plasmon frequency $\Omega_{p}$ of the electron gas in three dimensions (3DEG), which takes the universal value~\cite{pines52,pines89,giuliani05},
\begin{align}
\Omega_{p,{\rm 3deg}}^2 &= \frac{4\pi e^2 n}{\kappa m} . \label{eq:3deg} 
\end{align}
This result depends only on the band mass $m$, the electron charge $e$, the background dielectric constant $\kappa$, and the density $n$. Importantly, Eq.~\eqref{eq:3deg} does not involve the Planck quantum $\hbar$, i.e., it is a completely classical quantity~\cite{pines89}. In particular, the electron gas plasmon frequency is not affected by electron interactions or a magnetic field, hence plasmons do not provide insight in the equation of state of other quantum properties of the system. That Eq.~\eqref{eq:3deg} is exact for a charged Fermi liquid (regardless of any many-body approximation) follows from arguments for sum rules of the dielectric function $\varepsilon(\omega,{\bf q})$~\cite{pines89}:
\begin{align}
m_p({\bf q}) &= - \int_0^\infty d\omega \, \omega^p \, {\rm Im} [\varepsilon^{-1}(\omega,{\bf q})] , \label{eq:sumrules}
\end{align}
where the positive quantity $m_p$ denotes the energy-integrated sum rule with weight $\omega^p$. The dielectric function encodes the density response (and hence the collective mode spectrum) of the system and is related to the dynamic structure factor by the fluctuation-dissipation theorem. The imaginary part of the inverse dielectric function ${\rm Im} [\varepsilon^{-1}(\omega,{\bf q})]$ is known as the loss function since it is proportional to the differential cross section of inelastic x-ray or electron scattering off the system~\cite{ritchie57}. The loss function typically consists of a broad background formed by particle-hole excitations (which can describe either inter- or intraband transitions), with the plasmon showing up as a resonance~\cite{powell59,chiarello18}. The long-wavelength plasmon~\eqref{eq:3deg} is undamped and thus determined by the zeros of the dielectric function $\varepsilon(\omega,{\bf q})$. Assuming then that the loss function at long wavelengths is exhausted by the plasmon pole and neglecting other contributions (the so-called single-mode approximation), Eq.~\eqref{eq:3deg} follows using the $f$-sum rule $m_1({\bf q}) = \frac{2\pi^2 e^2 n}{m}$ and the perfect screening sum rule $m_{-1}({\bf q}) = \frac{\pi}{2} + {\cal O}(q^2)$, since in this case $\Omega_{p,{\rm 3deg}}^2 = m_1/m_{-1}$~\cite{pines89}. Indeed, it turns out that the single-mode approximation is exact at long wavelengths, i.e., the plasmon pole exhausts the sum rules~\eqref{eq:sumrules}, with other contributions such as particle-hole excitations suppressed due to phase-space restrictions and dielectric screening~\cite{pines89}. In particular, even interband excitations (for models with multiple bands) do not dominate the sum rules: In a spectral decomposition, the continuity equation relates interband terms at order ${\cal O}(q^0)$ (the same order as the plasmon pole) to matrix elements of the total current operator, i.e., the sum over velocities of all particles. However, such a contribution must vanish for translationally invariant states, because velocity and momentum are proportional in a parabolic system, and the total momentum generates translations of the whole state.

The above argument breaks down for Dirac materials, which thus admit a richer behavior of Dirac plasmons that is sensitive to many-body interactions. In Dirac materials, valence and conduction bands form linear band touching points, Dirac points, which are described by an effective continuum two-band Hamiltonian
\begin{align}
\hat{H} = \chi v_F \boldsymbol{\sigma} \cdot \hat{\bf p} , \label{eq:hamiltonian}
\end{align}
where $v_F$ is the Fermi velocity and $\hat{\bf p} = - i \hbar \boldsymbol{\nabla}$ is the momentum operator. Dirac cones appear in pairs of opposite chirality $\chi=\pm$. The Hamiltonian~\eqref{eq:hamiltonian} describes valence and conduction bands with linear dispersion $E_a({\bf p}) = s \chi v_F |{\bf p}|$, where we define the band index $s=\pm$. Different from an electron gas system, the velocity ${\bf v} = v_F \frac{{\bf p}}{|{\bf p}|}$ has constant magnitude $v_F$ and is not proportional to the momentum~\cite{throckmorton18}. Hence, although the effective description is translationally invariant, interband transitions will contribute at the same order as the plasmon mode. Indeed, they give a divergent contribution such that the sum rules in Eq.~\eqref{eq:sumrules} are not even well defined~\cite{sabio08,throckmorton15}. Hence, there are no universal sum rule constraints on the Dirac plasmon frequency. Note that this is a general result for Dirac semimetals and not an artifact of the low-energy model~\eqref{eq:hamiltonian}. While the $f$-sum rule $m_1(q)$ would be finite in a lattice model~\cite{throckmorton18}, it would not be exhausted by the plasmon pole, but interband transition would contribute at the same order, so that there is still no exact constraint on the plasmon frequency.

For an extrinsic system with doping density $n$, a calculation using the random phase approximation (RPA) that takes into account intraband excitations gives a plasmon frequency~\cite{dassarma09}
\begin{align}
\Omega_{p,{\rm 3dsm}}^2 &= \frac{4\pi v_F e^2 n}{\kappa \hbar k_F}, \label{eq:3dsm}
\end{align}
where $k_F = \bigl(\frac{3\pi^2 n}{g}\bigr)^{2/3}$ is the Fermi wavenumber and we allow for a multiplicity $g$ of Dirac cone pairs. Taking into account interband transitions gives an effective electronic contribution to the dielectric constant $\kappa$, which results in logarithmic corrections to the scaling of the plasmon frequency with density~\cite{lv13,throckmorton15} or temperature~\cite{hofmann15}. Intuitively, the intraband contribution~\eqref{eq:3dsm} is related to the electron gas plasmon~\eqref{eq:3deg} by a density-dependent effective Dirac mass $m = \hbar k_F/v_F$. Most importantly, since the plasmon frequency~\eqref{eq:3dsm} contains an explicit factor of $\hbar$, it is said to be a ``quantum plasmon'', which does not admit a classical limit $\hbar \to 0$~\cite{dassarma09}. Given this explicit dependence on $\hbar$ and the lack of universal sum rule constraints, an immediate question is if these quantum plasmons show more general properties compared to the electron gas and if they probe the equation of state.

\begin{figure}[t]
\raisebox{0.0cm}{\scalebox{0.9}{\includegraphics{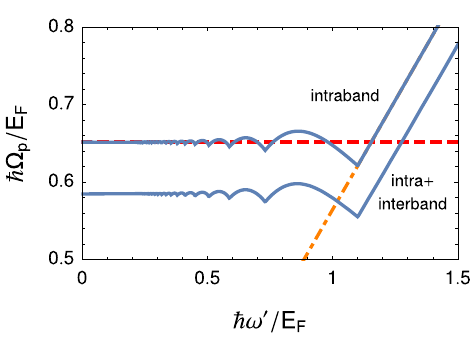}}\qquad}
\caption{Longitudinal plasmon frequency for $g\alpha=1$ as a function of $\hbar\omega'/E_F\sim\sqrt{B}$. The red dashed line indicates the zero-field plasmon frequency, and the dot-dashed orange line the high-field limit discussed in the main text.}
\label{fig:1}
\end{figure}

In this paper, we answer this question in the affirmative by showing that Dirac magnetoplasmons display quantum oscillations. We compute within the RPA the dielectric function of a Dirac semimetal in a constant external magnetic field and obtain an analytical result for the long-wavelength plasmon
\begin{align}
\Omega_{p,{\rm 3dsm}}^2(B) &= {\frac{4\pi v_F^2 e^2 n}{\kappa \mu(B)}} . \label{eq:longplasmon}
\end{align}
Distinct from the known electron gas plasmon, this frequency depends explicitly on the chemical potential $\mu$ and will thus show quantum oscillations of the de Haas-van Alphen type as the magnetic field $B$ is varied. To illustrate this, Fig.~\ref{fig:1} shows the plasmon frequency~\eqref{eq:longplasmon} as a continuous blue line as a function of the Dirac cylotron frequency $\omega' = \sqrt{2} v_F/\ell$, with $\ell = \sqrt{\hbar/eB}$ the magnetic length. The quantum oscillations are clearly visible. Intuitively, our result is related to the zero field case~\eqref{eq:3dsm} by replacing the Fermi energy $E_F = \hbar v_F k_F$ with the chemical potential. We include a full discussion of features of plasmons and the loss functions.

It turns out that the quantum oscillations in Eq.~\eqref{eq:longplasmon} are due to intraband transitions in the dispersing 3DSM Landau levels (LL). By contrast, magnetoplasmons in two-dimensional semimetals (2DSM) like graphene consist of interband transition between adjacent Landau levels. For comparison, we include a discussion of the 2DSM case in this paper. For collective modes in 3DSM that propagate at an angle to the external field, both mechanisms play out, which is discussed as well.

This paper is structured as follows: We will work on the level of the random phase approximation to determine the dynamic structure factor and the collective mode spectrum. For Dirac semimetals, the RPA is a very reliable and accurate many-body technique that represents the exact leading order for a large multiplicity of Dirac cones. To this end, Sec.~\ref{sec:II} discusses single-particle properties of Dirac semimetals (Sec.~\ref{sec:IIa}) in a magnetic field and introduces the RPA (Sec.~\ref{sec:IIb}). Results of this calculation are presented in~\ref{sec:III}, with a main focus on 3DSM in Sec.~\ref{sec:IIIa}. We present analytic results for the long-wavelength response and derive the central result for the longitudinal plasmon frequency, Eq.~\eqref{eq:longplasmon}. We also present a detailed discussion of the RPA loss function at all momenta and frequencies as a function of the magnetic field as well as the alignment between the field and the direction of the excitation. The section also contains a discussion of the 3DEG, which is the canonical model for interacting electrons used to discuss plasmons. The comparison with the 3DEG serves to highlight the markedly distinct and richer new behavior of Dirac plasmons. In Sec.~\ref{sec:IIIb}, we discuss plasmons in 2DSM as well as 2DEG as a function of a perpendicular magnetic field. The paper is concluded in Sec.~\ref{sec:conclusion}.

\section{Electromagnetic response and random phase approximation}\label{sec:II}

This section sets up the random phase approximation for Dirac materials in a constant magnetic field. We begin by summarizing the single-particle properties of Dirac particles in two and three dimensions in Sec.~\ref{sec:IIa}. The RPA calculation of the dielectric function and density response is presented in Sec.~\ref{sec:IIb}.

\subsection{Single-particle properties}\label{sec:IIa}

The Dirac Hamiltonian in an external magnetic field is obtained by substituting $\hat{\bf p} \to \hat{\bf p} - e \hat{\bf A}$ in Eq.~\eqref{eq:hamiltonian}, where we choose the Landau gauge for the vector potential $\langle {\bf r} | \hat{\bf A} | {\bf r} \rangle = (0,Bx,0)$ that describes a constant magnetic field in the $z$-direction. Eigenstates are given by~\cite{ashby13,ashby14,burkov15}
\begin{align}
|n, s, \chi, p_y, p_z\rangle &= \begin{pmatrix}u_{ns\chi}(p_z) |n, p_y, p_z\rangle \\ v_{ns\chi}(p_z) |n-1, p_y, p_z\rangle \end{pmatrix} \label{eq:eigenstate}
\end{align}
where $|np_yp_z\rangle$ is the single-particle eigenstate of the three-dimensional electron gas in a magnetic field,
\begin{align}
\langle {\bf r} |np_yp_z\rangle &= \frac{e^{i p_y y/\hbar + i p_z z/\hbar}}{\sqrt{L_yL_z}} \frac{1}{\sqrt{\ell}} \phi_n\Bigl(\frac{x + \ell^2 p_y/\hbar}{\ell}\Bigr) , \label{eq:wf}
\end{align} 
and $\phi_n$ is the dimensionless wave function of the harmonic oscillator in one dimension, $\phi_n(s) = \frac{e^{-s^2/2}}{\sqrt{2^n n!}} H_n(s)$, where $H_n$ is a Hermite polynomial. The coefficients in Eq.~\eqref{eq:eigenstate} are
\begin{align}
u_{ns\chi}(p_z) &= \begin{cases} 1 & n=0 \\ \sqrt{\frac{1}{2} + \frac{s \chi v_F p_z}{2 |E_{na}(p_z)|}} & n \neq 0 \end{cases} \label{eq:deff} , \\
v_{ns\chi}(p_z) &= \begin{cases} 0 & n=0 \\ s \chi \sqrt{\frac{1}{2} - \frac{s \chi v_F p_z}{2 |E_{na}(p_z)|}} & n \neq 0 \end{cases} . \label{eq:defg}
\end{align}
The eigenstates~\eqref{eq:eigenstate} have energy
\begin{align}
E_{ns\chi}(p_z) = \begin{cases} 
\chi v_F p_z & n=0 \\
s \chi \sqrt{(\hbar \omega')^2 n + (v_F p_z)^2} & n\neq 0
\end{cases} ,
\end{align}
where we introduce the Dirac cyclotron frequency $\omega' = \sqrt{2} v_F/\ell$ with $\ell = \sqrt{\hbar/eB}$ the magnetic length. Landau level (LL) states with $n=0$ have linear dispersion with a slope set by the chirality of the Weyl point. We will compare results with the three-dimensional electron gas, for which the single-particle spectrum is bounded from below with energy $E_n(p_z) = \hbar \omega_c (n+\tfrac{1}{2}) + \tfrac{p_z^2}{2m}$, where $\omega_c = \tfrac{\hbar}{ml_B^2} = \tfrac{eB}{m}$ is the 3DEG cyclotron frequency.

\begin{figure}[t]
\raisebox{0.0cm}{\scalebox{0.8}{\includegraphics{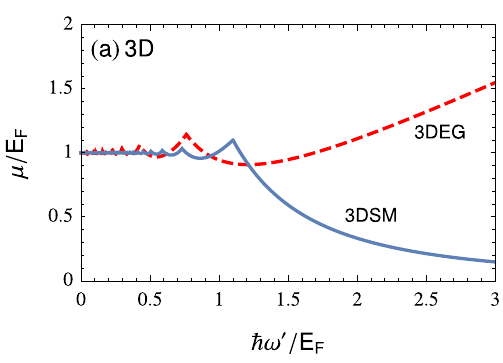}}\qquad}
\raisebox{0.0cm}{\scalebox{0.8}{\includegraphics{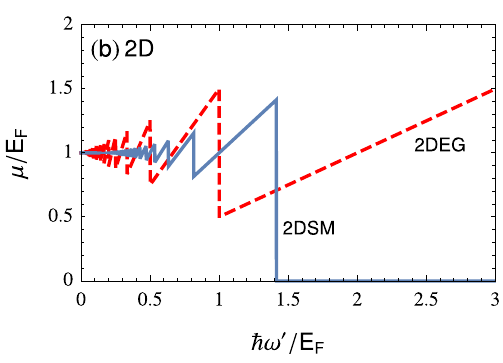}}\qquad}
\caption{Chemical potential at zero temperature as a function of magnetic field strength (continuous blue line) in (a) 3D and (b) 2D. For comparison, we show the chemical potential of a normal Fermi gas as a red dashed line. $\omega'$ denotes the respective cyclotron frequency of the semimetal and the electron gas.}
\label{fig:2}
\end{figure}

We determine the chemical potential (choosing $\mu>0$) requiring the doping density $n= \frac{g k_F^3}{3\pi^2}$ to be independent of the magnetic field,
\begin{align}
n &= \frac{g}{2 \pi^2 \ell^2} \biggl[\frac{\mu}{\hbar v_F} + 2 \sum_{n=1}^\infty K_n \, \Theta\bigl(\frac{\mu}{\hbar \omega'} - \sqrt{n}\bigr)\biggr] , \label{eq:den}
\end{align}
with $K_n=\frac{1}{\hbar v_F} \sqrt{\mu^2 - (\hbar\omega')^2 n}$. The first term in square brackets in Eq.~\eqref{eq:den} is the zeroth Landau level (0LL) contribution, and the prefactor accounts for the degeneracy of states $(2\pi\ell^2)^{-1}$ per unit area perpendicular to the magnetic field. The results of this calculation for the chemical potential as a function of the magnetic field is shown in Fig.~\ref{fig:2}(a) as a blue continuous line. In the low-field limit, using the Euler-MacLaurin formula in Eq.~\eqref{eq:den}, we find $\frac{\mu}{E_F} = 1$ as expected. In the opposite high-field limit only the zeroth Landau level contributes such that $\frac{\mu}{E_F} = \frac{4}{3} \bigl(\frac{E_F}{\hbar \omega'}\bigr)^2$, which vanishes with the inverse of the magnetic field. As the magnetic field changes, the chemical potential shows quantum oscillations with cusps at field values $\hbar \omega_n' = \sqrt{n} E_F$ whenever a Landau level is fully depopulated. For comparison, we include in Fig.~\ref{fig:2}(a) the chemical potential of the 3DEG as a red dashed line. The results are qualitatively similar, the noticeable difference being the different frequency of cyclotron oscillations as well as a different high-field behavior that arises from the field dependence of the 3DEG lowest Landau level. 

Note that the frequency of oscillations in Figs.~\ref{fig:1} and~\ref{fig:2} in a Dirac semimetal will receive corrections due to electron interactions~\cite{henriksen10,throckmorton18,sokolik17}. This is different for an electron gas, where the frequency of quantum oscillations is set by the cyclotron frequency, which is fixed at $\omega_c = eB/m$ and is not renormalized (this is Kohn's theorem~\cite{kohn61}). The failure of Kohn's theorem is linked to the specific form of the electron velocity, which is no longer proportional to the momentum~\cite{throckmorton18}, and is thus closely related to the failure of the single-mode approximation for the plasmon mode discussed in the introduction. Intuitively, it accounts for the fact that in a Dirac material, the collective motion of many electrons cannot be separated from the relative internal motion.

In two-dimensional Dirac semimetals (2DSM), there is a discrete single-particle Landau level spectrum with energy~\cite{roldan09,goerbig11},
\begin{align}
E_{na}(p_z) = \begin{cases} 
0 & n=0 \\
\chi s \hbar \omega' \sqrt{n} & n\neq 0
\end{cases} ,
\end{align}
with $s=\pm$ the band index and $\chi = \pm$ the chiral index and eigenstates,
\begin{align}
|n, s, \chi, p_y\rangle &= \begin{pmatrix}u_{n} |n, p_y\rangle \\ s \chi v_{n} |n-1, p_y\rangle \end{pmatrix} ,
\end{align}
where $u_{n} = \sqrt{(1+\delta_{n,0})/2}$, $v_{n} = \sqrt{(1-\delta_{n,0})/2}$, and $\langle {\bf r} |np_y\rangle = \frac{1}{\sqrt{L_y}} e^{i p_y y/\hbar} \frac{1}{\sqrt{l_B}} \phi_n\Bigl(\frac{x + l_B^2 p_y/\hbar}{l_B}\Bigr)$ is the single-particle eigenstate of the two-dimensional electron gas (2DEG). The density is
\begin{align}
n &= \frac{g k_F^2}{\pi} = \frac{k_F^2}{2 \pi} \times 2 g \biggl(\frac{\hbar \omega'}{E_F}\biggr)^2 \sum_{n=0}^{n_0} \nu_n ,
\end{align}
where the 0LL has occupation $\nu_0=1/2$. The result for the chemical potential as a function of magnetic field is shown in Fig.~\ref{fig:2}(b) as a blue continuous line. The chemical potential is fixed at the energy of the highest occupied Landau level, the energy of which increases linearly with the cyclotron energy $\hbar \omega'$. As the magnetic field increases, so does the degeneracy of states per Landau level. Above a critical field, the Landau level is depleted, upon which the chemical potential jumps discontinuously to the energy of the next-lowest Landau level. In the low-field limit, the chemical potential is equal to the Fermi energy. In the high-field limit, it is at the 0LL. For comparison, we include the corresponding result for the 2DEG as a red dashed line.

\subsection{Random phase approximation}\label{sec:IIb}

The collective plasmon mode is set by the zero of the dielectric function, which is given by
\begin{align}
\varepsilon(\omega, {\bf q}) &= 1 - V({\bf q}) \Pi(\omega, {\bf q}) , \label{eq:dielectric}
\end{align}
where $V({\bf q}) = 4\pi\hbar^2 e^2/q^2$ is the Coulomb interaction ($V({\bf q}) = 2\pi\hbar e^2/q$ in 2D) and $\Pi(\omega, {\bf q})$ is the screened density response that is irreducible with respect to the Coulomb interaction. In the RPA, we take $\Pi(\omega, {\bf q})$ as the noninteracting density response function:
\begin{align}
\Pi(\omega, {\bf q}) &= \frac{g}{2\pi\ell^2} \sum_{\mathclap{\substack{n,n'\\[-0.6ex]s,s'\\[-0.8ex]\chi,\chi'}}} \int \frac{dp_z}{2\pi\hbar} |F_{nn'}^{ss',\chi\chi'}(p_z, {\bf q})|^2 \nonumber \\
&\times\frac{f_0(E_{n's'\chi'}(p_z+q_z)) - f_0(E_{ns\chi}(p_z))}{E_{n's'\chi'}(p_z+q_z) - E_{ns\chi}(p_z) + \hbar \omega} \label{eq:polarization} ,
\end{align}
with $f_0$ the Fermi-Dirac distribution and $F_{nn'}^{ss',\chi\chi'}$ the density matrix element,
\begin{align}
&F_{nn'}^{ss',\chi\chi'}(p_z, {\bf q}) = u_{ns\chi}(p_z) u_{n's'\chi'}(p_z + q_z) f_{nn'}(q_x, q_y) \nonumber \\
&\quad+ v_{ns\chi}(p_z) v_{n's'\chi'}(p_z + q_z) f_{n-1,n'-1}(q_x, q_y) .
\end{align}
Here, $f_{nn'}(q_x, q_y) = \langle n | e^{i{\bf q}_\perp \cdot \hat{\bf r}} |n'\rangle$ (where ${\bf q}_\perp=(q_x,q_y)$) is the matrix element of the density operator between one-dimensional harmonic oscillator states $|n\rangle$. It can be expressed in closed analytical form in terms of Laguerre polynomials~\cite{giuliani05}. The matrix element $|F_{nn'}^{ss',\chi\chi'}(p_z, {\bf q})|^2$ only depends on the magnitude of ${\bf q}_\perp$, but the full response will still depend on the direction of propagation ${\bf q}$ with respect to the magnetic field in $z$-direction. We denote the angle between ${\bf q}$ and ${\bf B}$ by $\theta$. We will be mostly interested in the long-wavelength limit $q\to0$, where the plasmon is undamped.

In two dimensions, the density response takes the form~\cite{roldan09,goerbig11}
\begin{align}
\Pi(\omega, {\bf q}) &= \frac{g}{2 \pi \ell^2} \sum_{n,n'} \sum_{s,s'} \frac{f_0(E_{n's'}) - f_0(E_{ns})}{E_{n's'} - E_{ns} + \hbar \omega} |F_{nn'}^{ss'}({\bf q})|^2 \label{eq:Pi2D}
\end{align}
with
\begin{align}
F_{nn'}^{ss'}({\bf q}) &= u_{n} u_{n'} f_{nn'}({\bf q}) + s s' v_{n} v_{n'} f_{n-1,n'-1}({\bf q}) .
\end{align}

Note that for Dirac systems, the RPA is the leading order in an expansion in large orders of $g$, the multiplicity of Dirac cones. As such, it is a non-perturbative method valid for any value of the Dirac Coulomb interaction strength $\alpha = \frac{e^2}{\kappa \hbar v_F}$. Since $g$ is large for typical Dirac semimetals (for example, $g=12$ in TaAs~\cite{xu15,lv15,yan17} or pyrochlore iridates~\cite{wan11}), we expect the RPA to be quantitatively predictive. Note that the RPA provides an accurate description of many-body effects in graphene (for which $g$ is as low as $2$)~\cite{hofmann14}.

\section{Results}\label{sec:III}

In this section, we use RPA set up in Sec.~\ref{sec:IIb} to derive the plasmon mode and the loss function of Dirac semimetals. Section~\ref{sec:IIIa} presents results for the 3DSM. We derive an analytical expression for the longitudinal plasmon frequency and obtain the complete loss function as a function of frequency and momentum for different magnetic fields and propagation directions. Since the results of this paper are qualitatively distinct from the standard model of interacting electrons, the electron gas, we include a discussion of the 3DEG and compare with the 3DSM results. An additional point of comparison as already discussed in the introduction are 2DSM such as graphene, and Sec.~\ref{sec:IIIb} discusses both 2DSM and 2DEG plasmons and loss functions.

\begin{figure*}[t!]
\hspace{0.7cm}
$\, E_F/\hbar\omega'=0.9$
\hspace{1.3cm}
$\, E_F/\hbar\omega'=1.2$
\hspace{1.4cm}
$\, E_F/\hbar\omega'=2$
\hspace{1.5cm}
$\, E_F/\hbar\omega'=3$
\hspace{1.3cm}
$\, E_F/\hbar\omega'\to\infty$\\%
\vspace{0.3cm}
\hspace{.8cm}
\scalebox{0.2}{\includegraphics{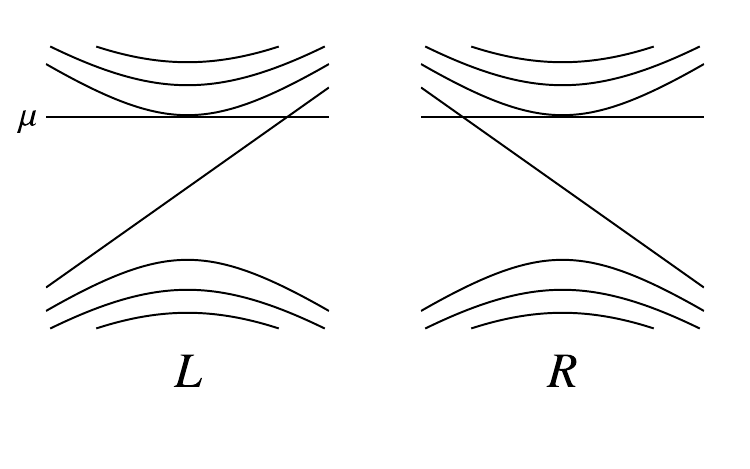}} \hspace{0.8cm}
\scalebox{0.2}{\includegraphics{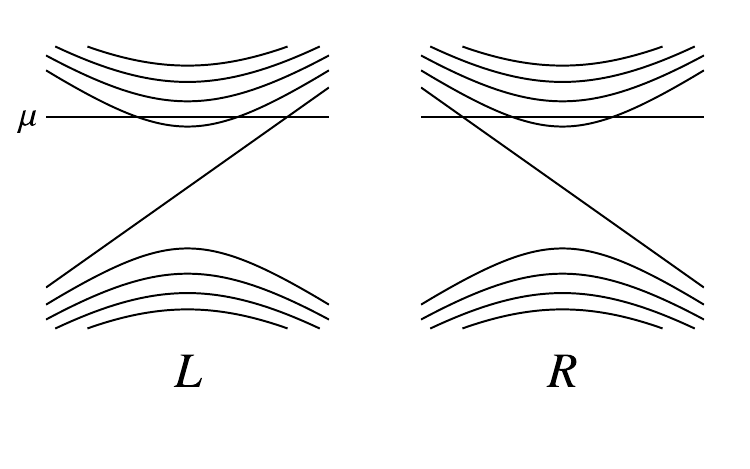}} \hspace{0.8cm}
\scalebox{0.2}{\includegraphics{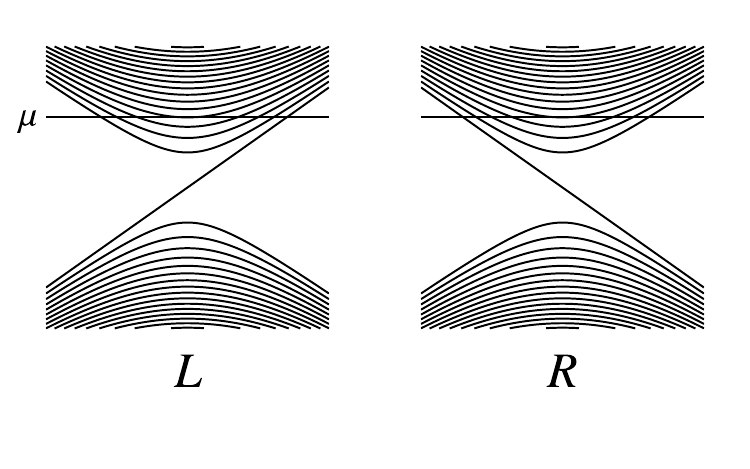}} \hspace{0.7cm}
\scalebox{0.2}{\includegraphics{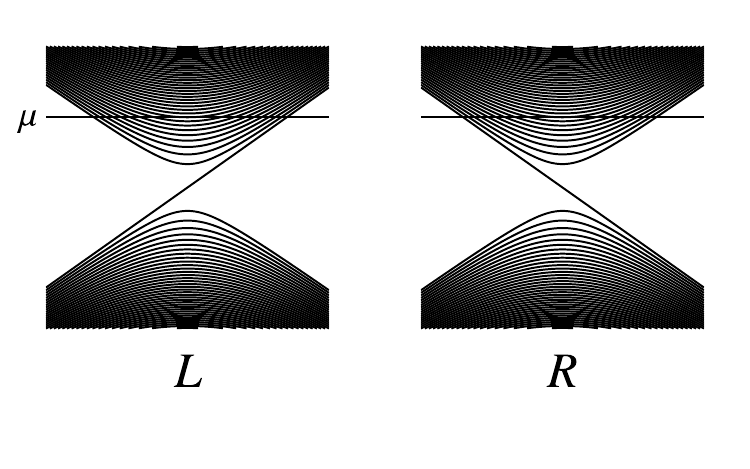}} \hspace{0.7cm}
\scalebox{0.2}{\includegraphics{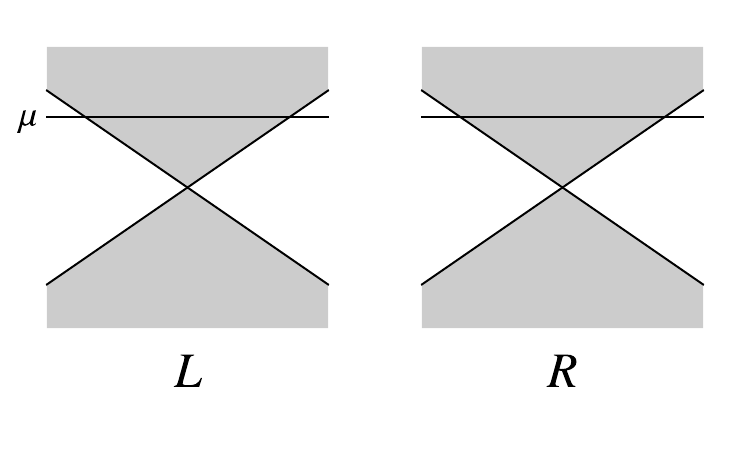}} \hspace{0.8cm}
\\%
\raisebox{1.3cm}{\rotatebox{90}{$\theta=0$}} 
\scalebox{0.39}{\includegraphics{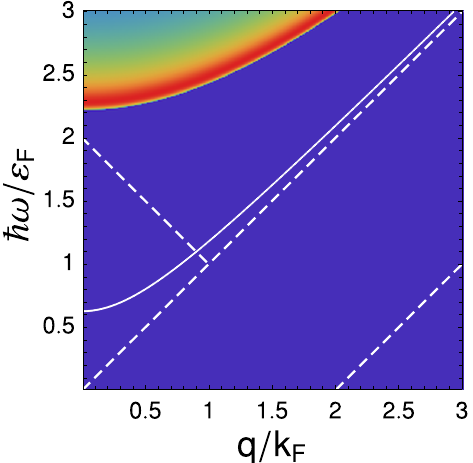}}\quad
\scalebox{0.39}{\includegraphics{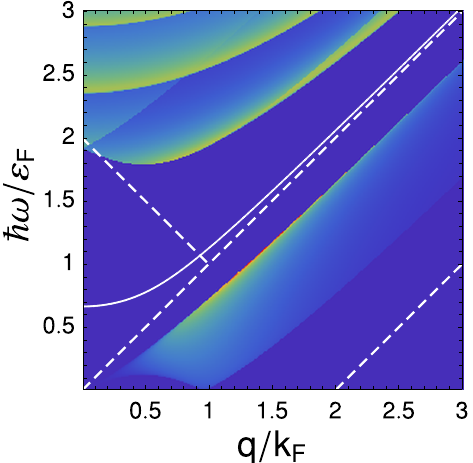}}\quad
\scalebox{0.39}{\includegraphics{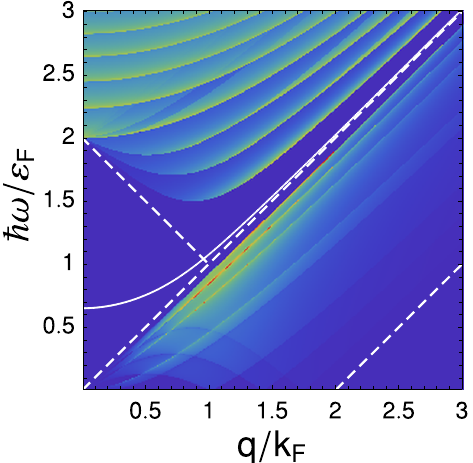}}\quad
\scalebox{0.39}{\includegraphics{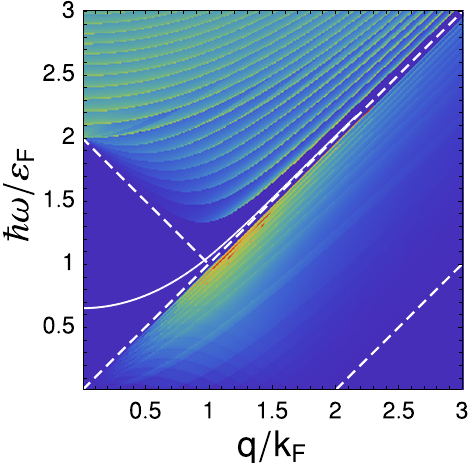}}\quad
\scalebox{0.39}{\includegraphics{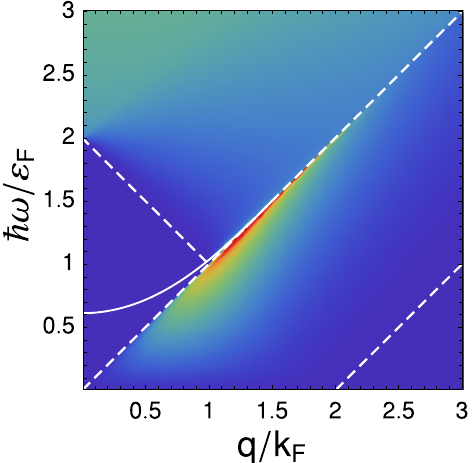}}\quad
\\%
\raisebox{1.2cm}{\rotatebox{90}{$\theta=\pi/4$}} 
\scalebox{0.39}{\includegraphics{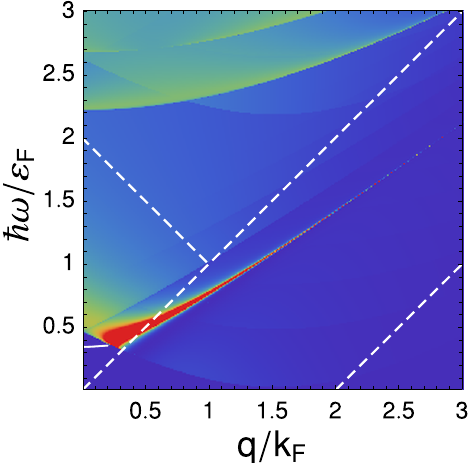}}\quad
\scalebox{0.39}{\includegraphics{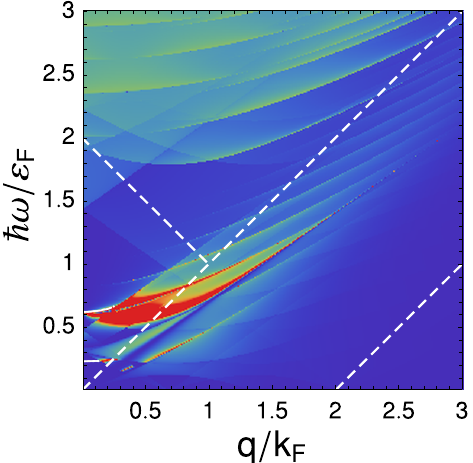}}\quad
\scalebox{0.39}{\includegraphics{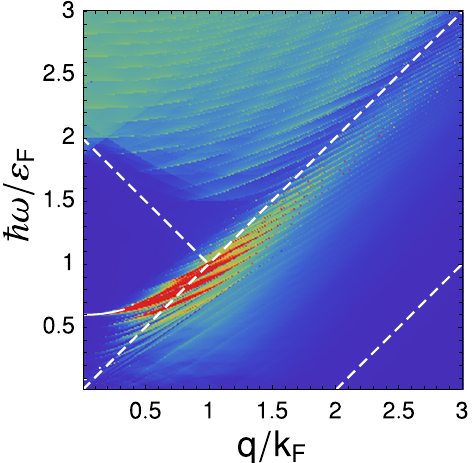}}\quad
\scalebox{0.39}{\includegraphics{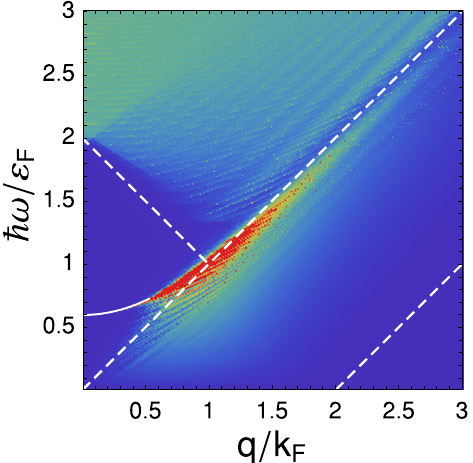}}\quad
\scalebox{0.39}{\includegraphics{fig3_25}}\quad
\\%
\raisebox{1.2cm}{\rotatebox{90}{$\theta=\pi/2$}} 
\scalebox{0.39}{\includegraphics{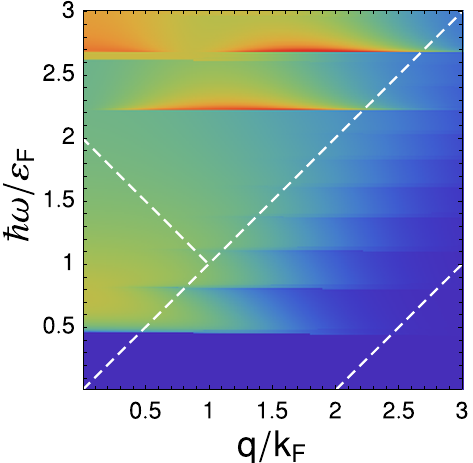}}\quad
\scalebox{0.39}{\includegraphics{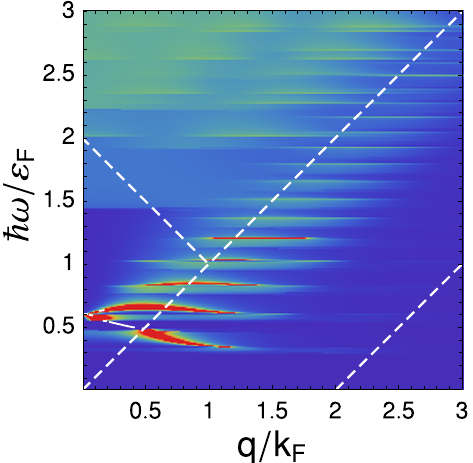}}\quad
\scalebox{0.39}{\includegraphics{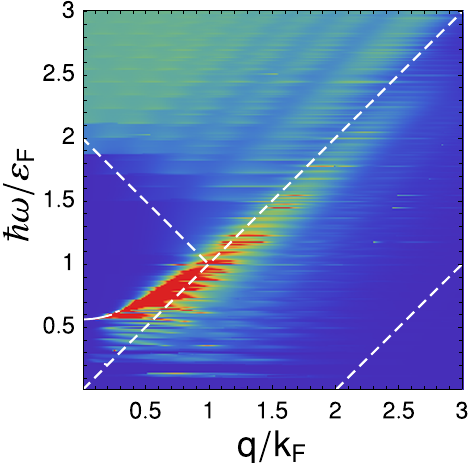}}\quad
\scalebox{0.39}{\includegraphics{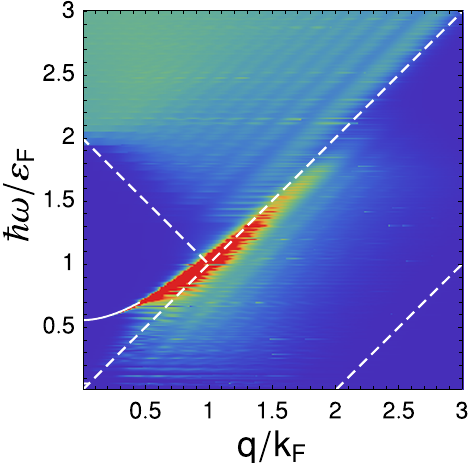}}\quad
\scalebox{0.39}{\includegraphics{fig3_25}}\quad
\\%
\caption{Loss function ${\rm Im} \, \varepsilon_{\rm RPA}^{-1}(\omega, q)$ of a Dirac semimetal with $g\alpha=1$ for various magnetic field strengths $E_F/\hbar\omega'=0.9,1.2,2,3$ and $\infty$ (left to right) and different angles $\theta$ of propagation ${\bf q}$ with respect to the magnetic field direction (top to bottom). At the very top we indicate the position of the chemical potential for both Dirac cones with opposite chirality. The rightmost panels correspond to the zero field limit and are the same for all angles. The continuous white line marks the plasmon dispersion and dashed white lines indicate the continuum particle-hole boundaries.
}
\label{fig:3}
\end{figure*}

\subsection{Three-dimensional semimetals}\label{sec:IIIa}

Consider first modes propagating along the magnetic field [i.e., ${\bf q} = (0,0,q_z)$]. In this case, $f_{nn'}(q_\perp) = \delta_{nn'} + {\cal O}(q_\perp^2)$, hence the long-wavelength limit of the polarization function is determined by transition between Landau levels with equal $n$ quantum number. These are intraband excitations within occupied conduction band levels, and interband excitations between opposite valence and conduction band levels. The contribution of intraband excitations (the ``Dirac plasma'') can be evaluated in closed analytical form. It consists of a zeroth Landau level part and a contribution from higher Landau levels, $\Pi(\omega, q_z) = \Pi^{\rm 0LL}(\omega, q_z) + \Pi^{\rm HLL}(\omega, q_z)$, with
\begin{align}
\Pi^{\rm 0LL}(\omega, q_z) &= \frac{v_F}{2\pi^2 \hbar \ell^2} \frac{q^2}{(\hbar \omega)^2} + {\cal O}(q^4) \label{eq:Pi0LL} \\
\Pi^{\rm HLL}(\omega, q_z) &= \frac{q^2}{(\hbar\omega)^2} \frac{v_F^2}{\mu} \biggl[n - \frac{1}{2 \pi^2 \ell^2} \frac{\mu}{\hbar v_F}\biggr] + {\cal O}(q^4) , \label{eq:PiHLL}
\end{align}
which is obtained from Eq.~\eqref{eq:polarization} expanding $f_0(E_{na}(p_z+q_z)) - f_0(E_{na}(p_z)) = \frac{\partial f_0(E_{na})}{\partial p_z} q_z$ and $E_{na}(p_z+q_z) - E_{na}(p_z) = \frac{\partial E_{na}}{\partial p_z} q_z$ and using Eq.~\eqref{eq:den}. The result~\eqref{eq:Pi0LL} for the zeroth Landau level agrees with~\cite{panfilov14}. Substituting this result in Eq.~\eqref{eq:dielectric}, we obtain the plasmon frequency~\eqref{eq:longplasmon}, which can be recast in the form
\begin{align}
\Omega_p^2 &= {\frac{4\pi v_F^2 e^2 n}{\kappa \mu}} = {\frac{4\pi \hbar v_F^3 \alpha n}{\mu}} 
= \frac{E_F^2}{\hbar^2} {\frac{4 g \alpha}{3\pi} \frac{E_F}{\mu}} . \label{eq:longplasmon2}
\end{align}
Figure~\ref{fig:1} shows the plasmon frequency for $g\alpha = 1$ as a function of magnetic field. The quantum oscillations are clearly visible. The red dashed line indicates the low-field limit of Eq.~\eqref{eq:3dsm}, $\lim_{B\to0} \Omega_p = \sqrt{4 g \alpha/3\pi} E_F/\hbar$. The orange dot-dashed line marks the high-field limit $\lim_{B\to\infty} \Omega_p = \omega' \sqrt{g \alpha/\pi}$, which agrees with the results of previous works on the magnetoplasmon mode for very large magnetic fields~\cite{son13,panfilov14,tolsma17,long18}. 

\begin{figure*}[t!]
\hspace{0.6cm}
$\, E_F/\hbar\omega_c=1.2$
\hspace{1.4cm}
$\, E_F/\hbar\omega_c=2$
\hspace{1.5cm}
$\, E_F/\hbar\omega_c=3$
\hspace{1.5cm}
$\, E_F/\hbar\omega_c=5$
\hspace{1.5cm}
$\, E_F/\hbar\omega_c\to\infty$\\%
\vspace{0.3cm}
\hspace{0.8cm}
\scalebox{0.1}{\includegraphics{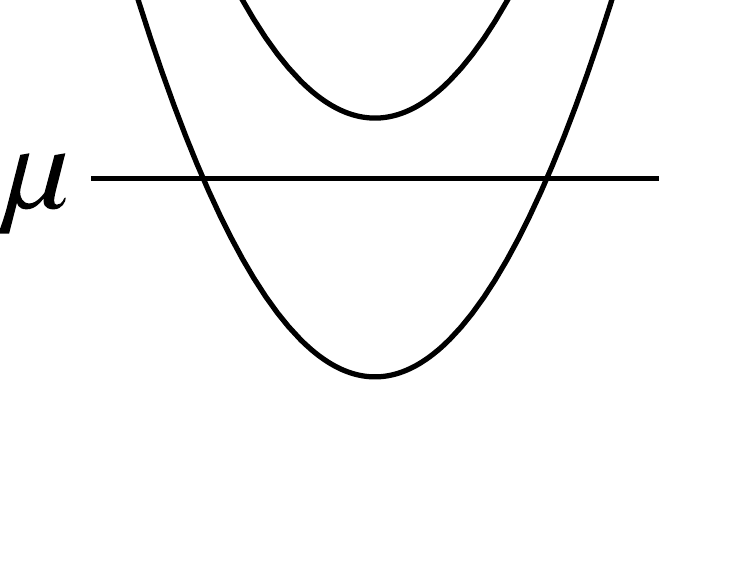}} \hspace{2.cm}
\scalebox{0.1}{\includegraphics{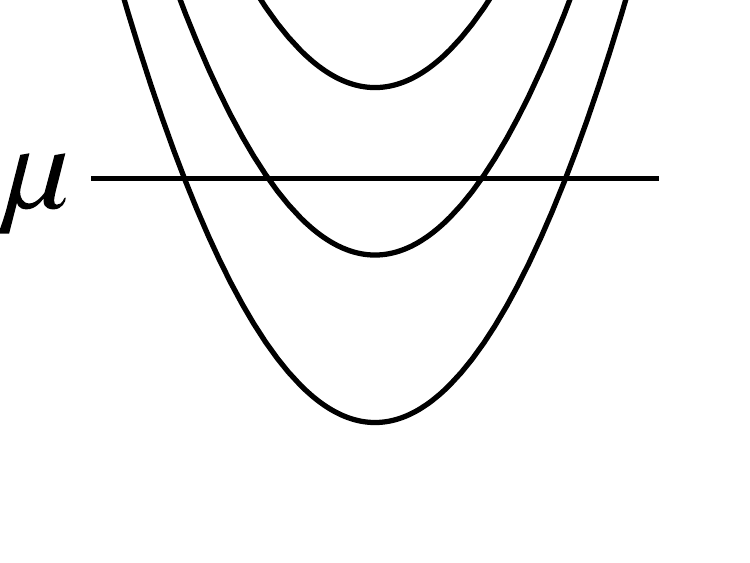}} \hspace{2.cm}
\scalebox{0.1}{\includegraphics{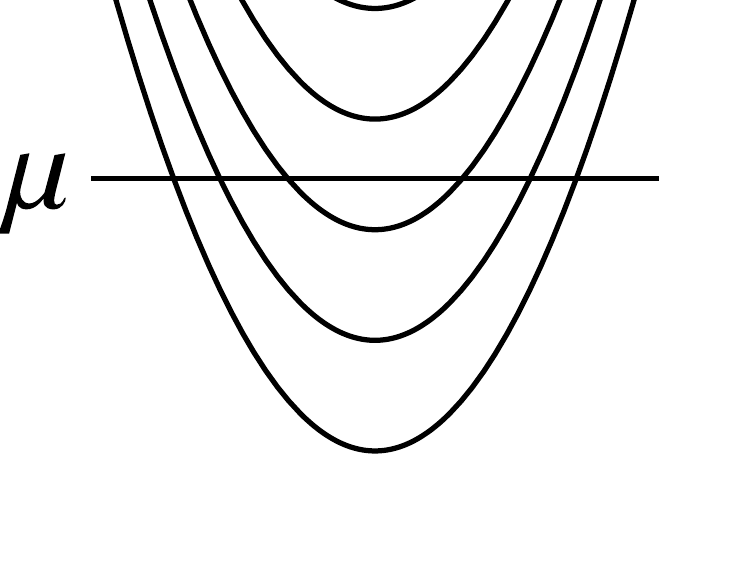}} \hspace{2.0cm}
\scalebox{0.1}{\includegraphics{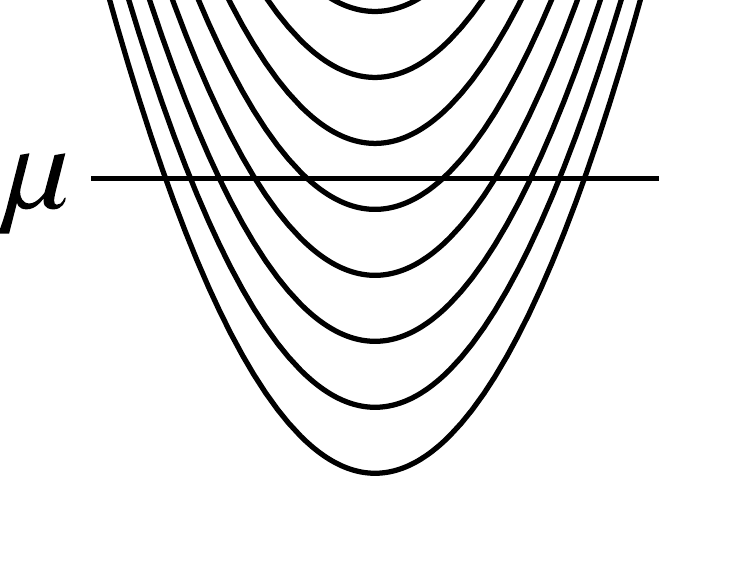}} \hspace{2.1cm}
\scalebox{0.1}{\includegraphics{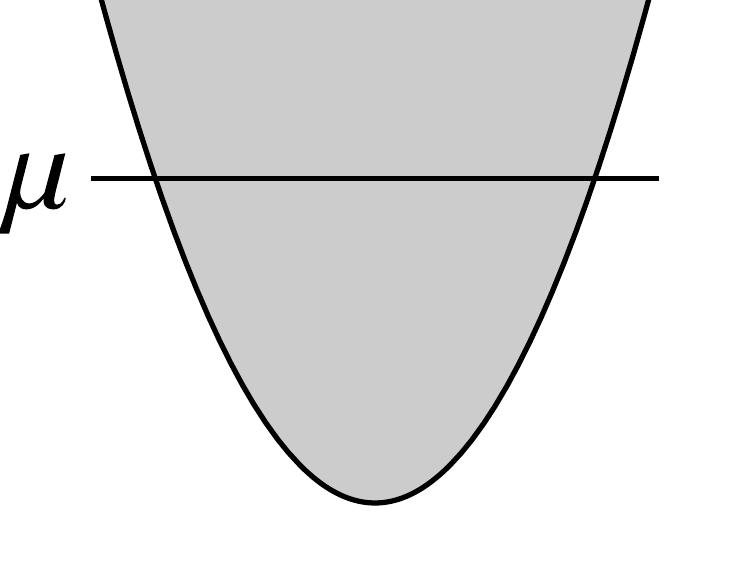}} \hspace{1.8cm}
\\%
\vspace{0.1cm}
\raisebox{1.3cm}{\rotatebox{90}{$\theta=0$}} 
\scalebox{0.25}{\includegraphics{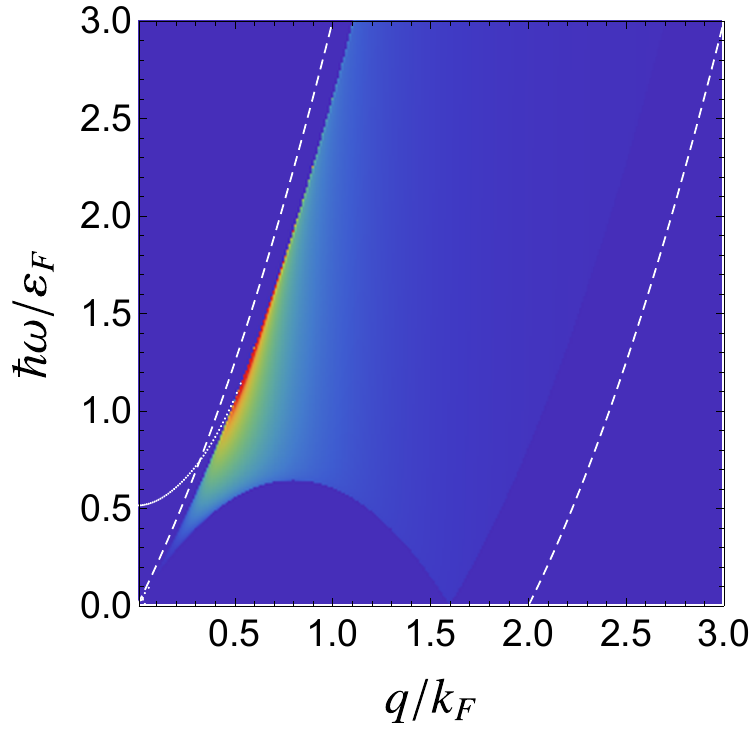}}\quad
\scalebox{0.25}{\includegraphics{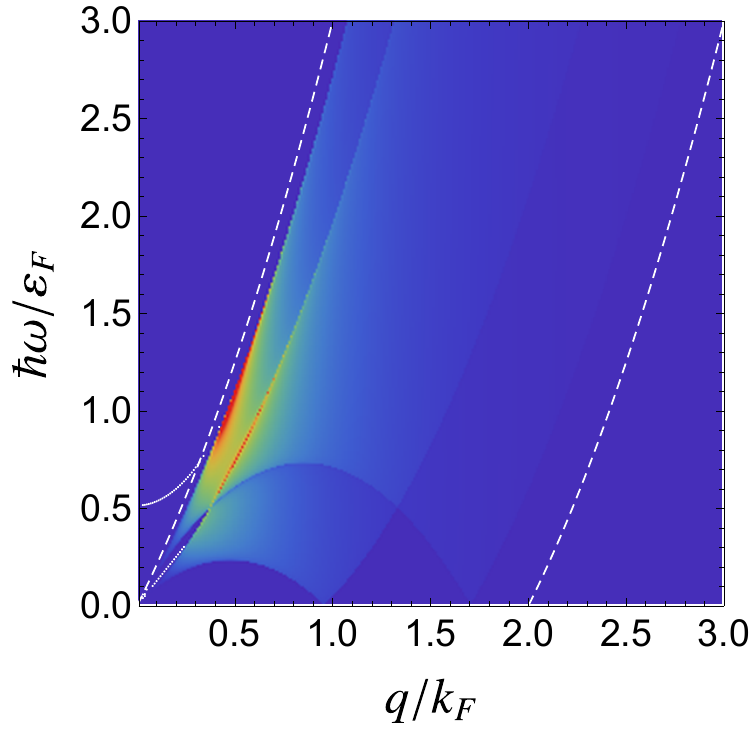}}\quad
\scalebox{0.25}{\includegraphics{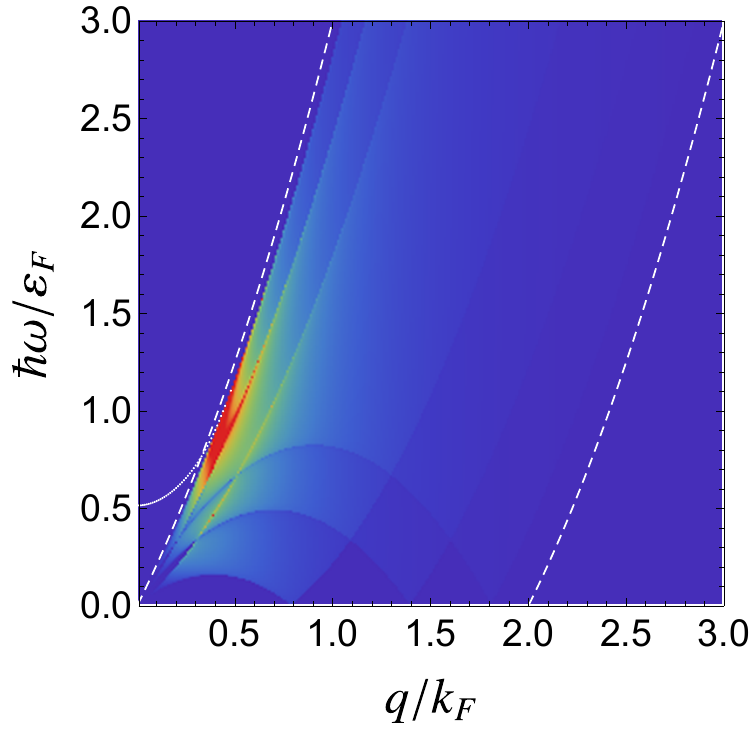}}\quad
\scalebox{0.25}{\includegraphics{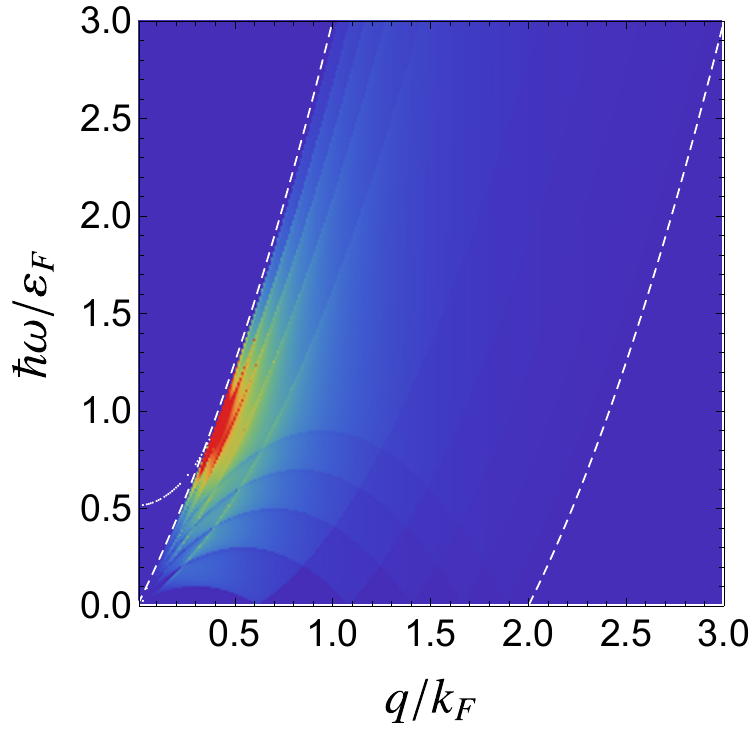}}\quad
\scalebox{0.25}{\includegraphics{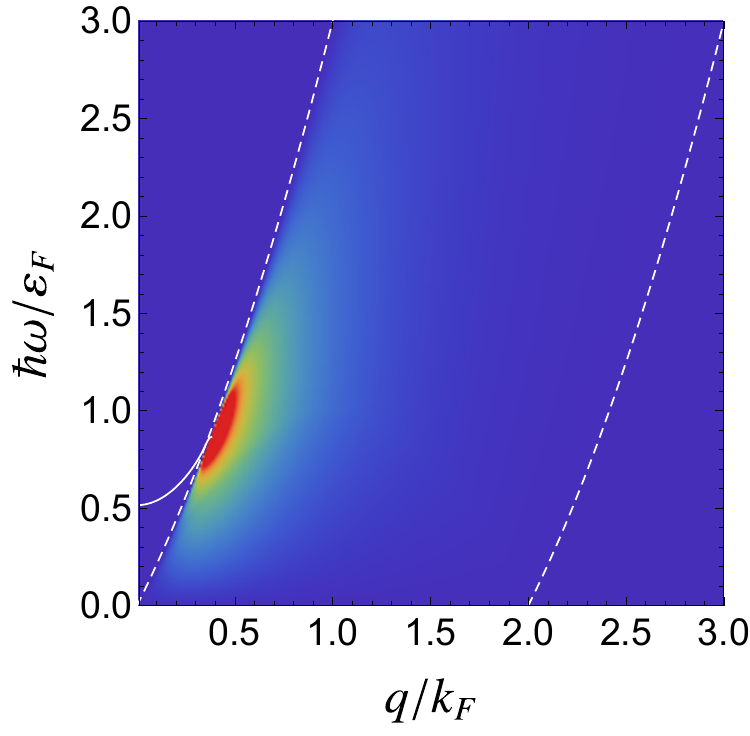}}\quad\\%
\raisebox{1.2cm}{\rotatebox{90}{$\theta=\pi/4$}} 
\scalebox{0.25}{\includegraphics{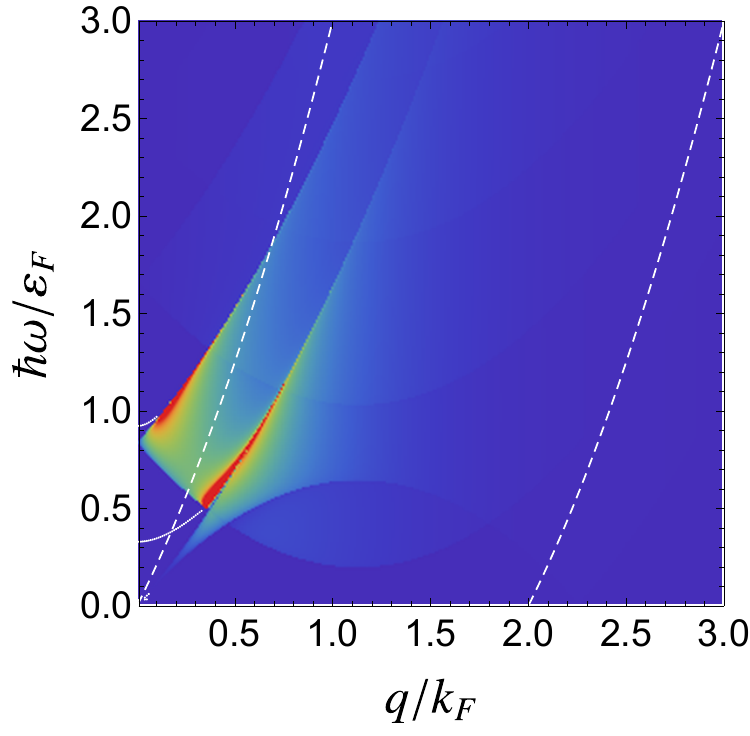}}\quad
\scalebox{0.25}{\includegraphics{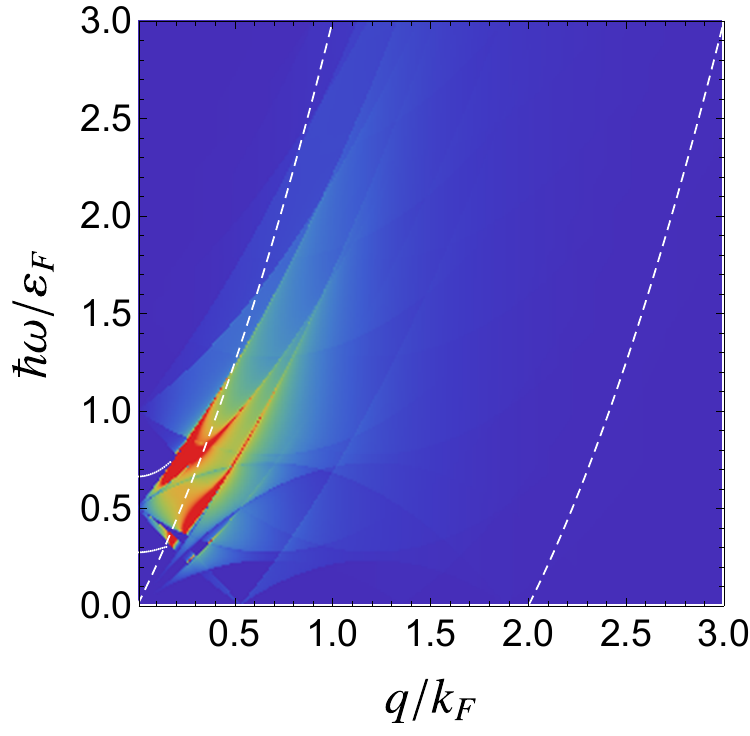}}\quad
\scalebox{0.25}{\includegraphics{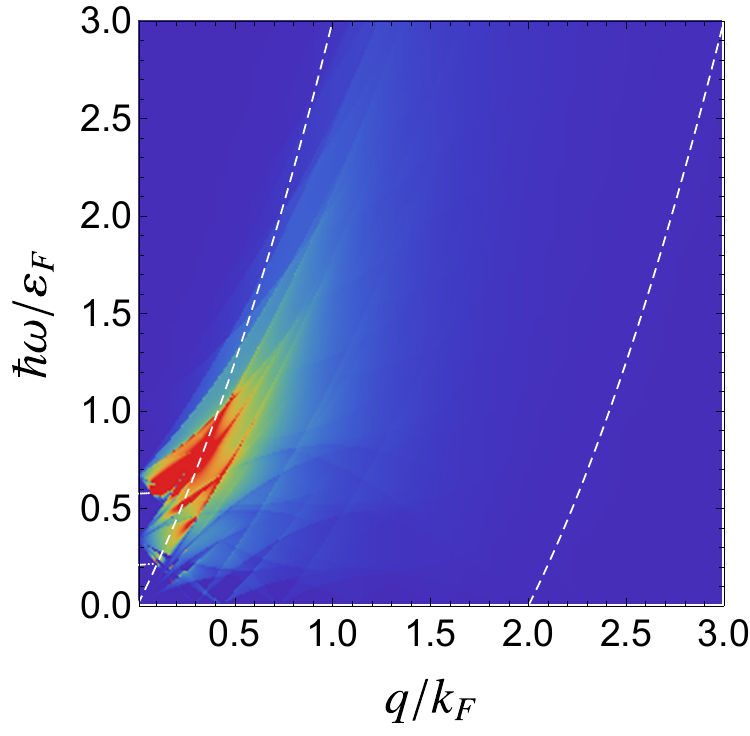}}\quad
\scalebox{0.25}{\includegraphics{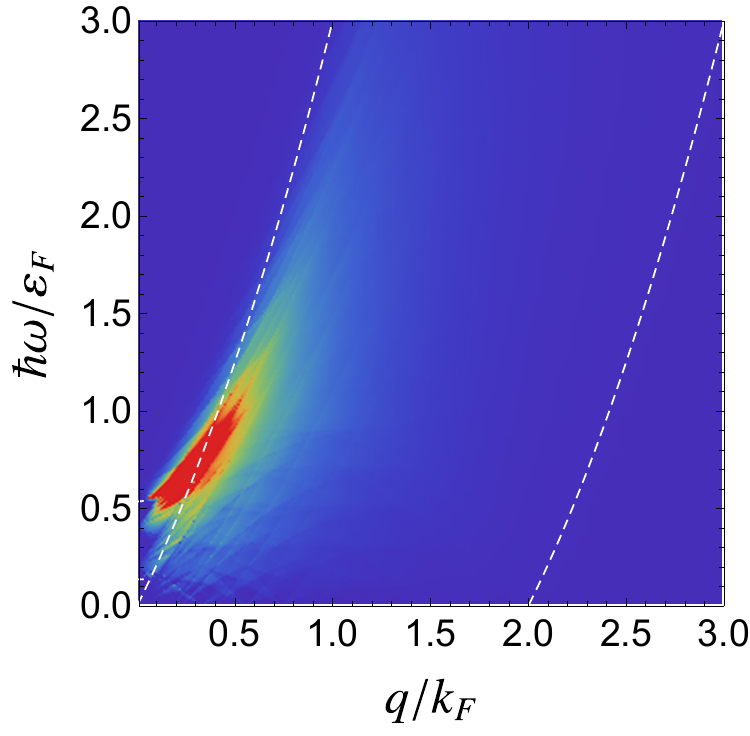}}\quad
\scalebox{0.25}{\includegraphics{fig4_25}}\quad\\%
\raisebox{1.2cm}{\rotatebox{90}{$\theta=\pi/2$}} 
\scalebox{0.25}{\includegraphics{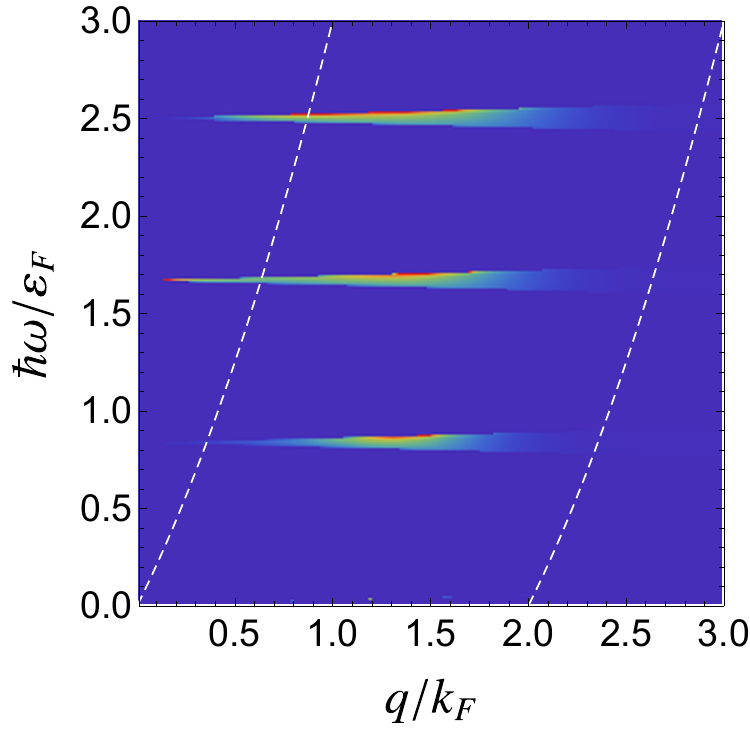}}\quad
\scalebox{0.25}{\includegraphics{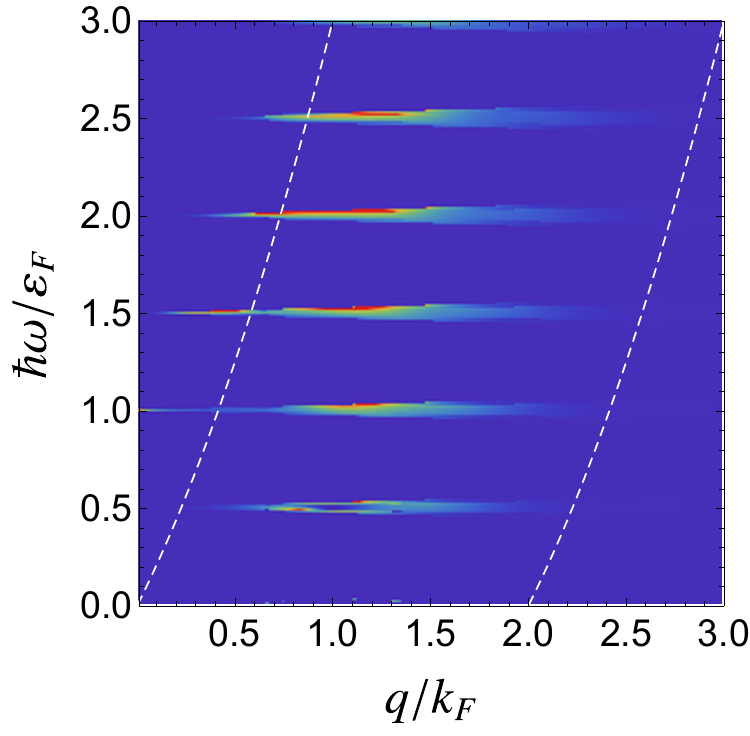}}\quad
\scalebox{0.25}{\includegraphics{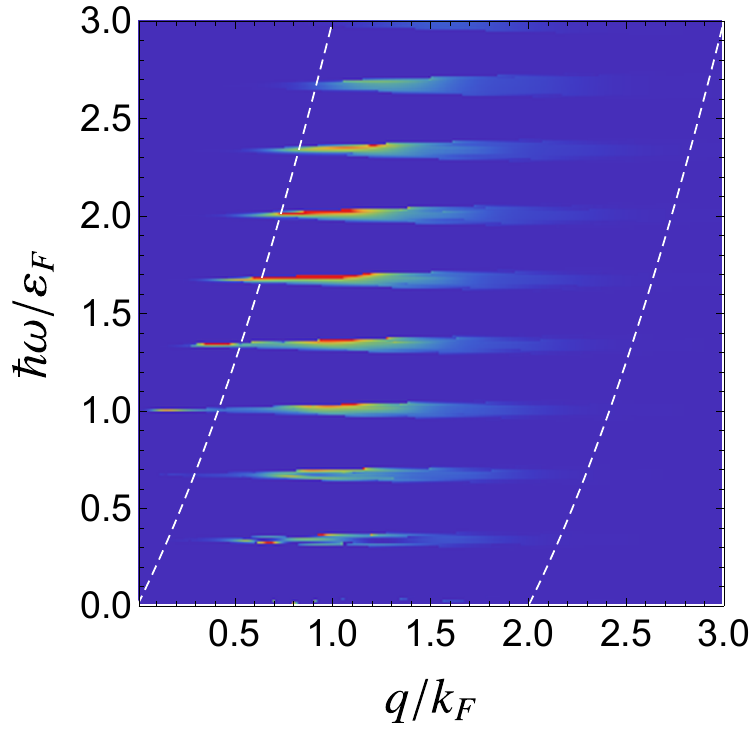}}\quad
\scalebox{0.25}{\includegraphics{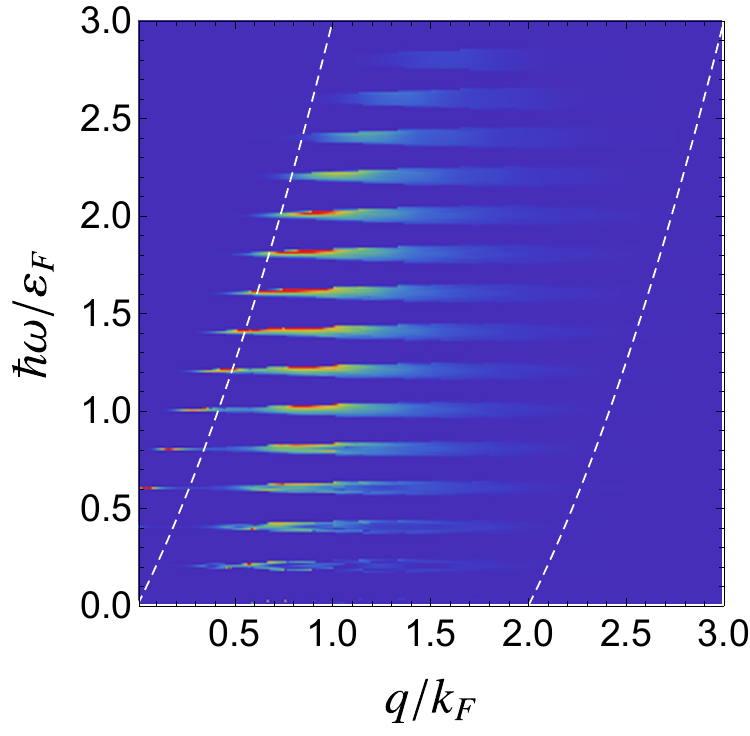}}\quad
\scalebox{0.25}{\includegraphics{fig4_25}}
\caption{Loss function of a 3DEG for various ratios $E_F/\hbar\omega_c$. $\theta$ denotes the angle of the propagation ${\bf q}$ with respect to the magnetic field direction. The rightmost panel is the same for all angles. The continuous white line marks the plasmon dispersion and dashed withe lines indicate the continuum particle-hole boundaries. As soon as the propagation in no longer parallel to the magnetic field, there are multiple plasmon modes and a tower of particle-hole continua (corresponding to inter-LL transitions) that coalesce at the Kohn mode frequency. }
\label{fig:4}
\end{figure*}

The crucial point about the results~\eqref{eq:longplasmon} and~\eqref{eq:longplasmon2} is that the Dirac plasmon mode depends on the equation of state through the chemical potential and thus shows quantum oscillations as the magnetic field changes. This is a main result of this work. By contrast, the plasmon frequency of the electron gas is constant and independent of the magnetic field~\cite{mermin64}, as required by the sum rule arguments discussed in the introduction. We stress that these quantum oscillations are a very general effect in that they are tied to the linear dispersion of Dirac or Weyl materials, they do not require anomalous electrodynamics. More broadly, the sensitivity of plasmon modes to many-body physics should be present for systems with a general non-parabolic dispersion $E(p) \sim p^{2+\beta}$ ($\beta \neq 0$) or $E(p) \sim \frac{p^2}{2m^*} + \gamma p^4$. 

For comparison, Fig.~\ref{fig:1} also shows the plasmon frequency for the full dielectric function including interband transitions. In the long-wavelength limit, only interband excitations between Landau levels with equal quantum number $n$ contribute to the polarization function~\eqref{eq:polarization}, such that the excitation energy is large compared to the plasmon frequency. As can be seen from Eq.~\eqref{eq:polarization}, interband transitions will thus contribute at order ${\cal O}(q^2/\omega^0)$ to the polarization function [compared to ${\cal O}(q^2/\omega^2)$ for intraband terms, Eqs.~\eqref{eq:Pi0LL} and~\eqref{eq:PiHLL}]. Hence, from Eq.~\eqref{eq:dielectric}, they provide an effective electronic contribution to the dielectric constant $\kappa$. This lowers the value of plasmon frequency compared to the value for the Dirac plasma. This result is in excellent agreement with the full calculation presented in Fig.~\ref{fig:1}. In order to compare calculations performed at different magnetic fields, we choose a magnetic-field dependent cutoff such that the density of electrons in the valence band remains unchanged (i.e., the chemical potential of the intrinsic system remains at the Dirac point). This illustrates that interband transitions do not qualitatively affect the quantum oscillations.

Note that the quantum oscillations in the magnetoplasmon mode presented in Fig.~\ref{fig:1} are not affected by internodal scattering~\cite{jenkins16,tolsma17}. Internodal scattering corresponds to a momentum transfer between different Dirac points that is typically much larger than the Fermi momentum. It does not affect the long-wavelength physics.

We now turn to a full discussion of the loss function ${\rm Im} [\varepsilon^{-1}(\omega, {\bf q})]$. Figure~\ref{fig:3} shows the RPA loss function for five different values of magnetic field (left to right) $E_F/\hbar \omega' = 0.9,1.2,2,5$ and $\infty$ (the latter is the zero-field limit) and three different angles (top to bottom) $\theta = 0,\pi/8$, and $\pi/4$ between the direction of the wave number and the magnetic field. At the very top, we show the single-particle spectrum of a pair of Dirac cones and indicate the position of the chemical potential for comparison. Where undamped, the plasmon mode is shown by a continuous white line. As is apparent from the figure, the loss functions cross over to the zero-field limit (shown in the right-most panel, which is independent of the magnetic field)~\cite{lv13}. Dashed white lines indicate the particle-hole boundary of the zero-field system. These boundaries follow from energy- and momentum conservation and describe the kinematic threshold for particle-hole excitations. For intraband transitions, the particle-hole continuum exists for ${\rm max}(0,\hbar v_F q - 2 \varepsilon_F) < \hbar \omega < \hbar v_F q$, and interband transitions exist for $\hbar \omega > {\rm max}(2 \varepsilon_F - \hbar v_F q, \hbar v_F q)$. The interband contribution to the loss function of the Dirac system at $\omega > q$ is non-zero at long wavelengths, which leads to the divergence of the sum rules discussed earlier.

The first row with $\theta = 0$ corresponds to the case discussed previously where the direction of propagation is aligned with the magnetic field. Only transitions between Landau levels with equal quantum number $n$ contribute to the loss function. There is an intraband contribution stemming from transition within a single Landau level and interband contributions from transitions between opposite valence band and conduction band Landau levels. The particle-hole continuum formed by intra-level transitions in nonzero Landau levels is characteristic of a one-dimensional system, where zero-frequency excitation at finite momentum are only possible if the two Fermi points are connected~\cite{giuliani05}. Particle-hole excitations in the zeroth Landau level do not form a continuum, but are linear $\hbar \omega = \hbar v_F q$ due to the linear dispersion of this Landau level. An important results of our calculations is that the plasmon is undamped even at higher momenta and should hence dominate the loss function even at large momentum transfer.

When the magnetic field is tilted with respect to the direction of propagation, transitions between Landau levels with any quantum number are permitted. In particular, the plasmon mode merges with the particle-hole continuum and is Landau damped at larger wave numbers. At larger magnetic fields, the plasmon hybridizes and we numerically find two separate collective modes, which is qualitatively similar as for the electron gas~\cite{mermin64}. It is important to note, however, that the Dirac magnetoplasmons show quantum oscillations even in a tiled magnetic field, whereas the electron gas plasmons never show such oscillations.

\begin{figure*}[t!]
\hspace{0.7cm}
$\, E_F/\hbar\omega'=0.9$
\hspace{1.3cm}
$\, E_F/\hbar\omega'=1.2$
\hspace{1.4cm}
$\, E_F/\hbar\omega'=2$
\hspace{1.5cm}
$\, E_F/\hbar\omega'=3$
\hspace{1.3cm}
$\, E_F/\hbar\omega'\to\infty$\\%
\vspace{0.2cm}
\raisebox{1.2cm}{\rotatebox{90}{$\rm 2DSM$}} \hspace{0.03cm}
\scalebox{0.25}{\includegraphics{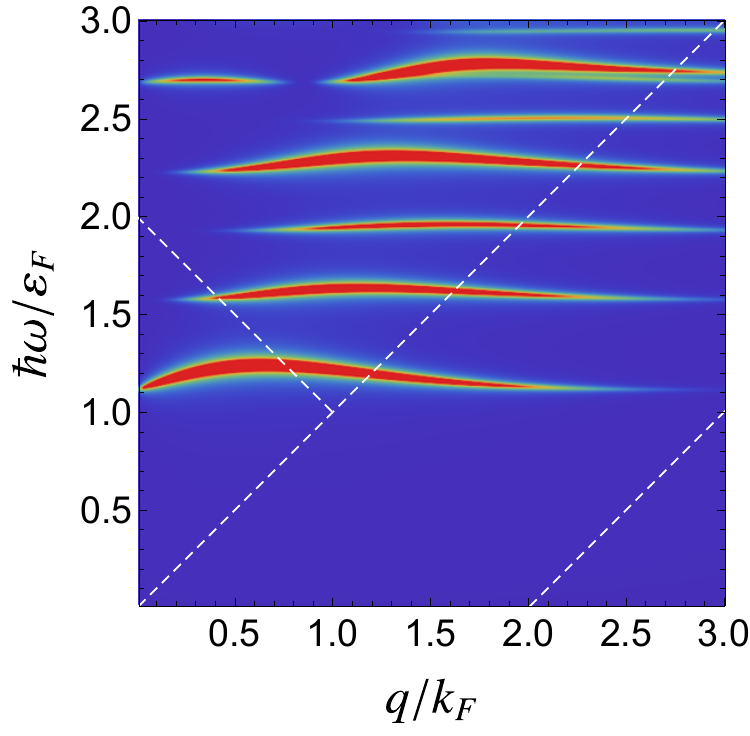}}\hspace{0.18cm}
\scalebox{0.25}{\includegraphics{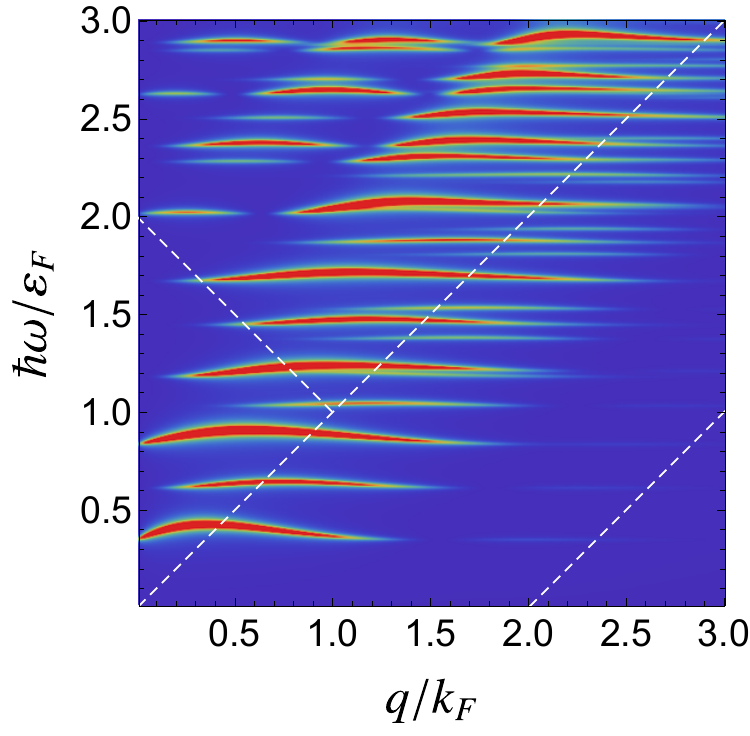}}\hspace{0.18cm}
\scalebox{0.25}{\includegraphics{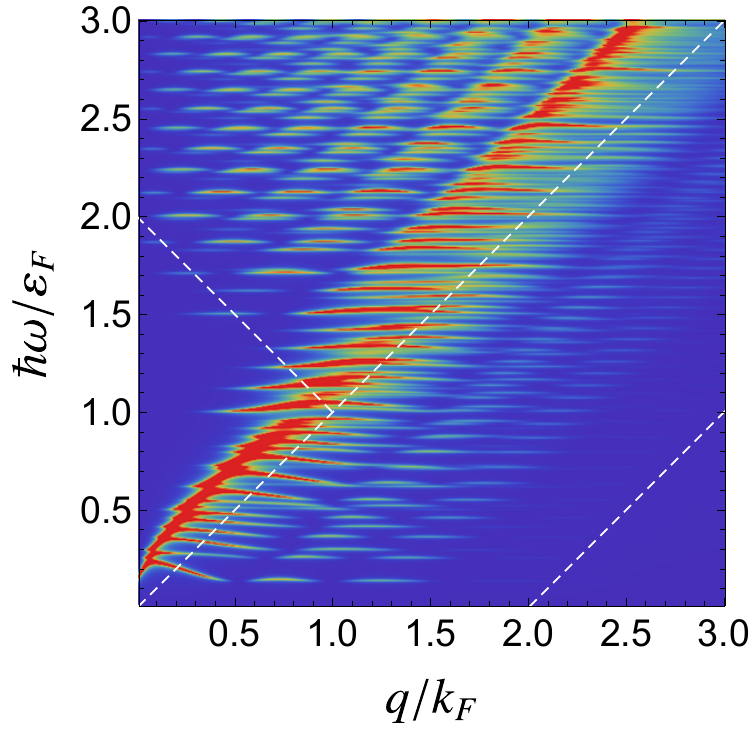}}\hspace{0.18cm}
\scalebox{0.25}{\includegraphics{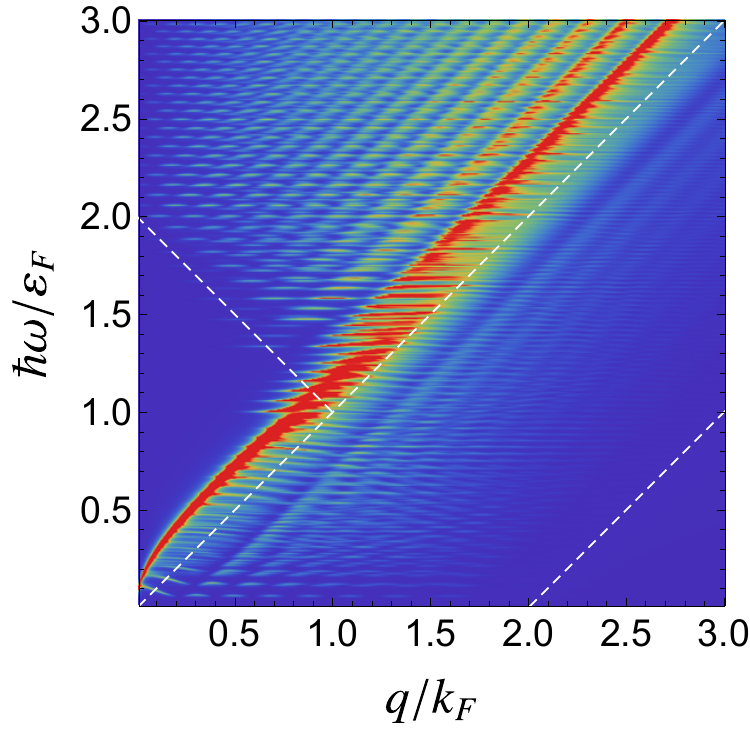}}\hspace{0.168cm}
\scalebox{0.25}{\includegraphics{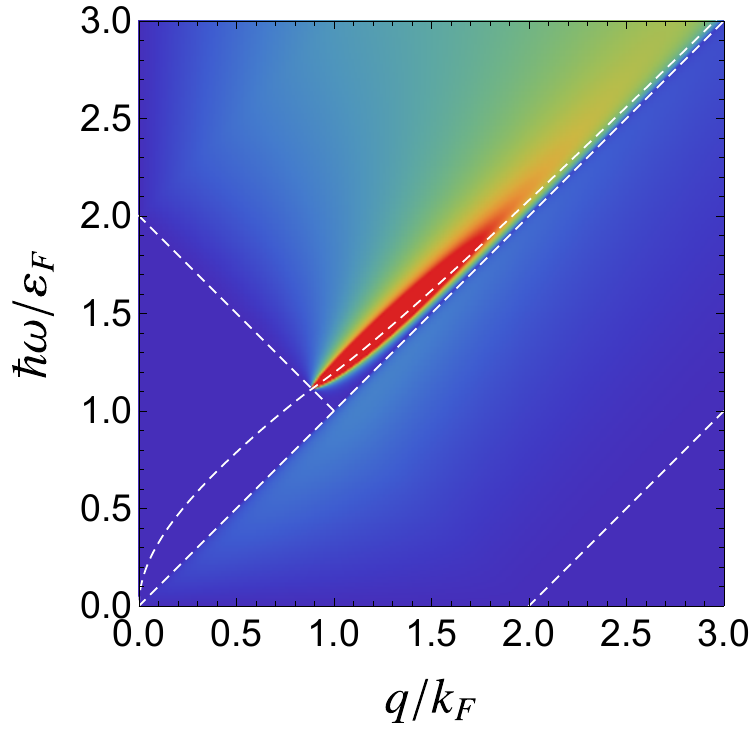}}\hspace{0.85cm}\\%
\vspace{0.5cm}
\hspace{0.7cm}
$\, E_F/\hbar\omega_c=1.2$
\hspace{1.3cm}
$\, E_F/\hbar\omega_c=2$
\hspace{1.4cm}
$\, E_F/\hbar\omega_c=3$
\hspace{1.5cm}
$\, E_F/\hbar\omega_c=5$
\hspace{1.3cm}
$\, E_F/\hbar\omega_c\to\infty$\\%
\vspace{0.2cm}
\raisebox{1.2cm}{\rotatebox{90}{$\rm 2DEG$}}  \hspace{0.03cm}
\scalebox{0.25}{\includegraphics{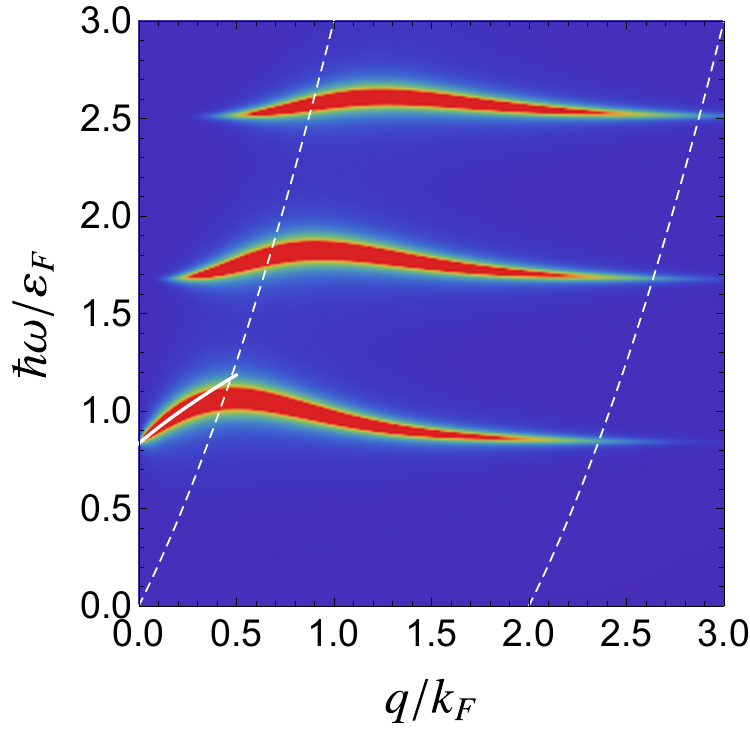}}\hspace{0.18cm}
\scalebox{0.25}{\includegraphics{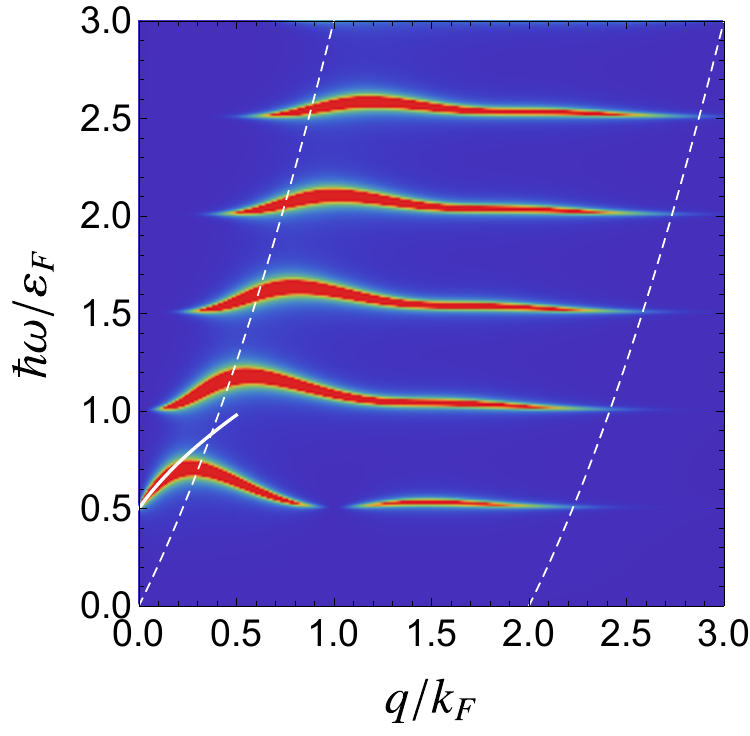}}\hspace{0.18cm}
\scalebox{0.25}{\includegraphics{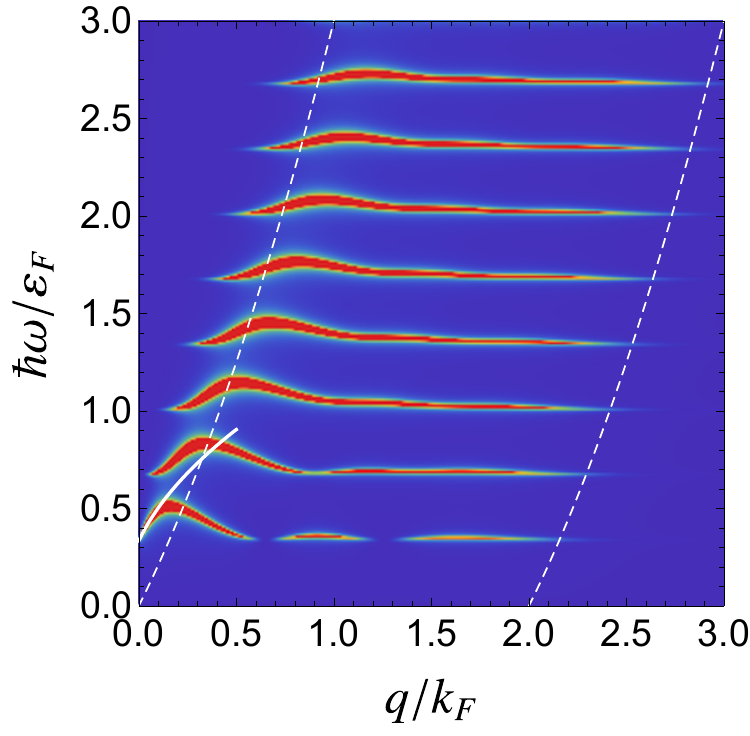}}\hspace{0.18cm}
\scalebox{0.25}{\includegraphics{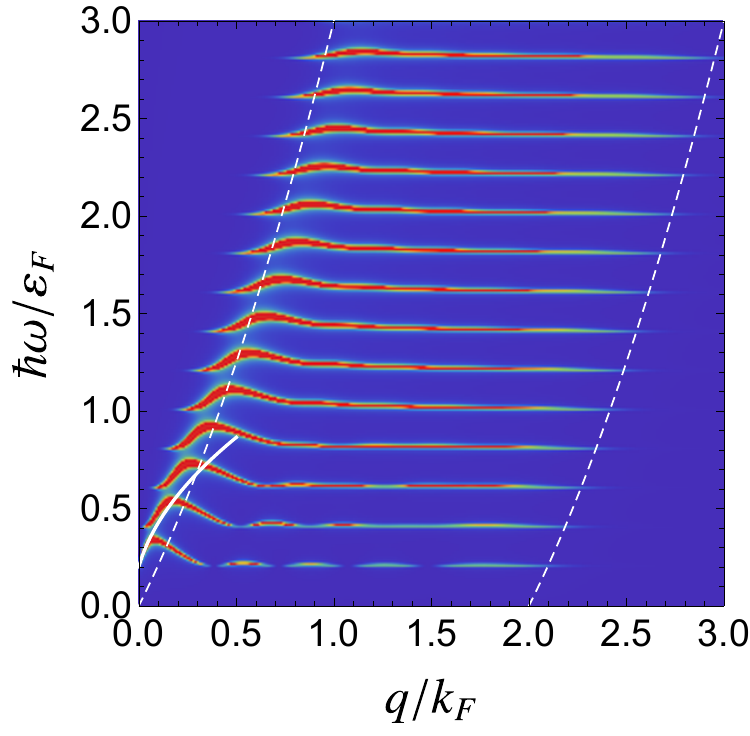}}\hspace{0.18cm}
\scalebox{0.25}{\includegraphics{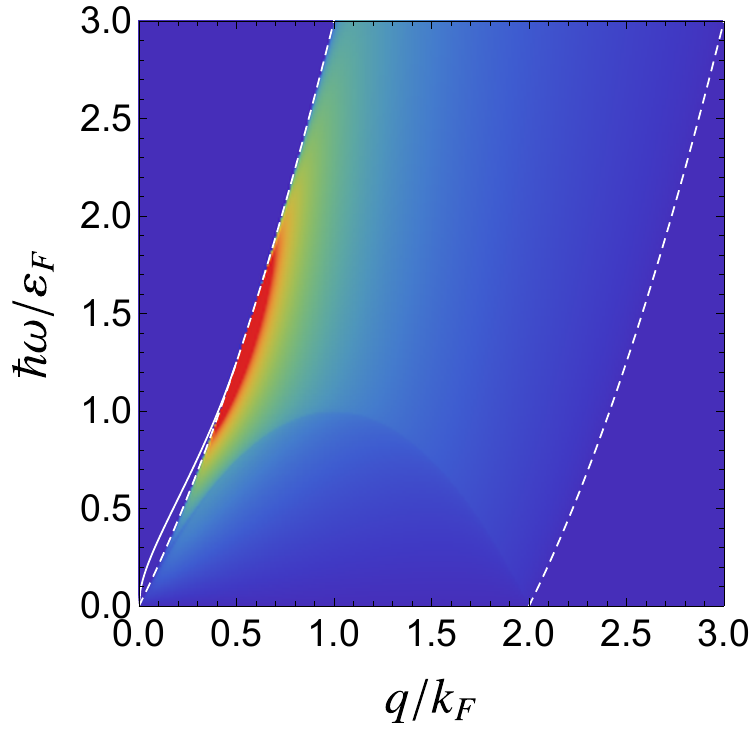}}\quad\\%
\caption{Top panels: Loss function of a 2DSM for various ratios $E_F/\hbar\omega_c$. The white lines indicate the magnetoplasmon. Bottom panels: Loss function of a 2DEG with $r_s = 0.5$ for various ratios $E_F/\hbar\omega_c$. The white lines indicate the magnetoplasmon.
}
\label{fig:5}
\end{figure*}

For comparison, Fig.~\ref{fig:4} shows the RPA loss function of the 3DEG for five different values of magnetic field (left to right) $E_F/\hbar \omega_c = 1.2,2,3,5$ and $\infty$ (the zero-field limit) and three different angles (top to bottom) $\theta = 0,\pi/8$, and $\pi/4$ between the direction of the wave number and the magnetic field. The collective plasmon mode is indicated by a white continuous line where undamped. The white dashed lines show the intraband particle-hole boundary of the zero-field system at frequencies $\hbar \omega_\pm = {\rm max}(0,\frac{\hbar^2q^2}{2m} \pm \hbar v_F q)$. The longitudinal plasmon mode (top panels) is independent of the magnetic field and fixed at the zero-field value, Eq.~\eqref{eq:3deg}. For excitations with a transverse momentum component with respect to the magnetic field ($\theta \neq 0$), there are two collective long-wavelength modes with angle-dependent frequency~\cite{mermin64}
\begin{align}
\omega_\pm^2 &= \frac{\Omega_p^2 + \omega_c^2}{2} \pm \frac{1}{2} \sqrt{(\Omega_p^2 + \omega_c^2)^2 - (2 \Omega_p \omega_c \cos \theta)^2} .
\end{align}
The splitting is apparent in the collective mode dispersion in the second row of Fig.~\ref{fig:4}. The splitting can be understood as the hybridization of the zero-field bulk plasmon mode and the cyclotron motion~\cite{mermin64}. Note that while these modes have a magnetic field dependence through the cyclotron frequency, it is trivial and does not show quantum oscillations or contains information about the equation of state.

\subsection{Two-dimensional semimetals}\label{sec:IIIb}

The quantum oscillations for longitudinal Dirac magnetoplasmons are due to intraband transitions in the dispersing 3DSM Landau level. By contrast, in 2DSM with a perpendicular magnetic field, the plasmon mode gaps out where the gap corresponds to interband transitions between different graphene Landau levels. For comparison, this section presents a discussion of the 2DSM and 2DEG magnetoplasmon mode. There is extensive previous literature on graphene magneoplasmons~\cite{roldan09,goerbig11}, which consider Landau levels with integer filling.

First, consider the long-wavelength limit. To order ${\cal O}(q^2)$, only terms with $n' = n\pm1$ will contribute in the polarization function~\eqref{eq:Pi2D}. Denote the index of the partially occupied LL by $m$ and consider $m\geq0$. To leading order in $q$, there are two magnetoplasmon modes, which are given by:
\begin{align}
&\frac{(\hbar\omega_1)^2}{(\hbar\omega')^2} = (\sqrt{m+1} - \sqrt{m})^2 \nonumber \\
&\quad+ \sqrt{2} \alpha (\sqrt{m+1} - \sqrt{m}) \frac{|F_{m,m+1}^{++}({\bf q})|^2}{l_B q/\hbar} \nu_m \\
&\frac{(\hbar\omega_2)^2}{(\hbar\omega')^2} = (\sqrt{m} - \sqrt{m-1})^2 \nonumber \\
&\quad + \sqrt{2} \alpha (\sqrt{m} - \sqrt{m-1}) \frac{|F_{m-1,m}^{++}({\bf q})|^2}{l_B q/\hbar} (1-\nu_m) .
\end{align}
The residue of the first mode is proportional to the filling fraction of the highest occupied Landau level $\nu_m$, and the residue of the second mode is proportional to $1-\nu_m$. Hence, there is a splitting of the long-wavelength magnetopolasmon for fractional filling. This splitting vanishes for a fully occupied LL, for which only the first mode has nonzero weight. This result is different from the electron gas, where there is no splitting of the plasmon mode for fractional filling owing to the equal energy spacing between LL. While the magnetoplasmon excitation in graphene systems has been considered before, previous studies consider fully-filled Landau levels~\cite{roldan09,goerbig11}, where the mode splitting of the long-wavelength plasmon discussed here is not apparent. Note that this behavior is also distinct from the 3DSM, and while there is a discontinuous jump in the plasmon frequency for certain magnetic fields, there is no quantum oscillation.

Results for the full 2DSM loss function are shown in Fig.~\ref{fig:5}(a) for five values of the magnetic field $E_F/\hbar \omega_c = 0.9,1.2,2,3$ and $\infty$. As before, the collective mode dispersion is indicated by the white continuous line, and the particle-hole zero-field boundary is shown as white dashed lines. The splitting in the long-wavelength limit is clearly visible in Fig.~\ref{fig:4}, most notably for $E_F/\hbar \omega'=1.2$. It is absent for the first panel with $E_F/\hbar \omega'=0.9$, where the chemical potential is in the zeroth LL, for which the excitations from occupied to empty bands have the same energy. In the low-field limit, the excitation gap vanishes, and the plasmon dispersion at long wavelength with the square root of the momentum.

Again, we compare with the corresponding results for the 2DEG. Figure~\ref{fig:5}(b) shows the loss function of the 2DEG for five values of the magnetic field $E_F/\hbar \omega_c = 1.2,2,3,5$ and $\infty$. The notation is the same as in Fig.~\ref{fig:5}(a). The long-wavelength RPA plasmon dispersion is obtained as
\begin{align}
\hbar \Omega_p &= \sqrt{(\hbar \omega_c)^2 + \frac{2 \pi e^2 n q}{m}} .
\end{align}
The second term is the zero-field plasmon mode. The magnetic field-dependence is trivial in the form of an offset at long wavelengths, which corresponds to direct interband excitations. Since the electron gas Landau levels have equal energy-spacing, there is no splitting of the collective mode as noted above.

\section{Conclusion}\label{sec:conclusion}

In conclusion, we have discussed collective magnetoplasmon excitations in Dirac semimetals. The main result is that Dirac plasmons show quantum oscillations, and are thus an important new probe of many-body physics in semimetals. Our findings are in contrast to the canonical system of many-body theory, the electron gas with parabolic bands, where the plasmon is a purely classical quantity that is not sensitive to the many-body physics. The unusual properties of Dirac plasmon provide a direct experimental signature of Dirac semimetals. With first experiments on three-dimensional Dirac plasmons appearing recently~\cite{sushkov15,chen15,jenkins16,xu16,chanana18,chiarello18}, the results of this paper should be accessible in current experiments. Going forward, the dependence on the magnetic field provides a novel way to tailor the Dirac plasmon frequency.

\begin{acknowledgements}
I thank Nigel Cooper for discussions. This work is supported by Peterhouse, Cambridge.
\end{acknowledgements}

\bibliography{bib}

\end{document}